\documentclass[letterpaper,titlepage,12pt]{article}
\usepackage[colorlinks=true,urlcolor=blue,citecolor=magenta]{hyperref}
\usepackage{amssymb,amsmath,amsfonts}
\usepackage{epsfig}
\usepackage{epsfig}
\usepackage{graphicx}
\usepackage{epstopdf}
\usepackage{caption}
\usepackage{subcaption}
\usepackage{amscd}
\usepackage{amsthm}
\usepackage{latexsym}
\usepackage{amsbsy}
\usepackage{bbm}
\usepackage[english]{babel}
\usepackage{psfrag}
\usepackage{tabularx}

\allowdisplaybreaks

\def\clock{{\count0=\time
           \divide\count0 60
           \ifnum\count0<10 0\fi\the\count0
           \multiply\count0 -60 \advance\count0 \time
           :\ifnum\count0<10 0\fi \the\count0
         }}
\newcommand{\timestamp}{{\small\vbox{\hbox{\tt\jobname.tex}
\hbox{\the\day/\the\month/\the\year, \clock}}}}

\numberwithin{equation}{section}

\begin{document}

\begin{titlepage}
\rightline{\vbox{   \phantom{ghost} }}
%
%
 \vskip 1.4 cm
\centerline{\LARGE \bf Holographic Reconstruction of   }
\vspace{.2cm}
\centerline{\LARGE \bf 3D Flat Space-Time}

\vskip 1.5cm

\centerline{\large {{\bf Jelle Hartong}}}

\vskip .8cm

\begin{center}

\sl Physique Th\'eorique et Math\'ematique and International Solvay Institutes,\\
Universit\'e Libre de Bruxelles, C.P. 231, 1050 Brussels, Belgium.
\vskip 0.4cm

\end{center}
\vskip 0.6cm

\vskip .8cm \centerline{\bf Abstract} \vskip 0.2cm \noindent

We study asymptotically flat space-times in 3 dimensions for Einstein gravity near future null infinity and show that the boundary is described by Carrollian geometry. This is used to add sources to the BMS gauge corresponding to a non-trivial boundary metric in the sense of Carrollian geometry. We then solve the Einstein equations in a derivative expansion and derive a general set of equations that take the form of Ward identities. Next, it is shown that there is a well-posed variational problem at future null infinity without the need to add any boundary term. By varying the on-shell action with respect to the metric data of the boundary Carrollian geometry we are able to define a boundary energy-momentum tensor at future null infinity. We show that its diffeomorphism Ward identity is compatible with Einstein's equations. There is another Ward identity that states that the energy flux vanishes. It is this fact that is responsible for the enhancement of global symmetries to the full BMS$_3$ algebra when we are dealing with constant boundary sources. Using a notion of generalized conformal boundary Killing vector we can construct all conserved BMS$_3$ currents from the boundary energy-momentum tensor.

\end{titlepage}

\pagestyle{empty}
\small
\tableofcontents

\normalsize
\newpage
\pagestyle{plain}
\setcounter{page}{1}


\section{Introduction}

Over the last few years quite some progress has been made in developing holographic tools for the study of various non-AdS holographic dualities. For example interesting progress has been made in the holographic description of Lifshitz, Schr\"odinger, warped AdS$_3$ and 3-dimensional flat space-times. Among these the approach to Lifshitz holography is the one that will be inspirational to the methods developed here for asymptotically flat space-times. 

It was shown in \cite{Christensen:2013lma,Christensen:2013rfa,Hartong:2014oma,Bergshoeff:2014uea,Hartong:2015wxa} that by employing boundary conditions in vielbein formalism for asymptotically Lifshitz space-times one can obtain a general covariant understanding of the boundary geometry which turned out to be torsional Newton--Cartan geometry (TNC) the details of which were worked out in the work on Lifshitz holography and independently in the context of field theory \cite{Geracie:2014nka,Jensen:2014aia,Hartong:2014pma,Jensen:2014wha,Hartong:2015wxa,Geracie:2015dea} and geometry \cite{Andringa:2010it,Bergshoeff:2014uea,Bekaert:2014bwa} including Ho\v rava--Lifshitz gravity \cite{Hartong:2015zia} and Newton--Cartan supergravity \cite{Andringa:2013mma,Bergshoeff:2015uaa,Bergshoeff:2015ija}. Carrollian geometry is a close cousin of TNC geometry that will play a very prominent role in this work. In fact there is a duality between these two concepts \cite{Duval:2014uoa,Bekaert:2015xua,Hartong:2015xda} that in 1+1 dimensions becomes an equivalence (see section \ref{subsec:2D} as well as \cite{Hofman:2014loa}). The observation that the boundary of an asymptotically Lifshitz space-time is described by torsional Newton--Cartan geometry allows one to define a boundary energy-momentum tensor and classify using geometrical tools the class of dual field theories that we are dealing with. The purpose of this paper is to show that the main ideas of \cite{Christensen:2013lma,Christensen:2013rfa,Hartong:2014oma,Bergshoeff:2014uea,Hartong:2015wxa} can also be applied to an entirely different class of space-times, namely 3D asymptotically flat space-times. 

A lot of progress has been made on flat space holography by viewing flat space-time as the large radius limit of AdS space-time \cite{Bagchi:2009my,Bagchi:2010zz,Barnich:2010eb,Barnich:2012aw,Krishnan:2013wta}. However this approach only works in 3 bulk dimensions whereas the really interesting case is in 4 bulk dimensions especially in regards to black hole space-times. Further it is esthetically not so attractive to have to resort to AdS computations to understand flat space. The large radius limit works for pure AdS giving rise to Minkowski space-time. Furthermore in 3 bulk dimensions all solutions are orbifolds of Minkowski/AdS. Hence the limit maps both the equations of motion as well as the entire solution space, but this does not happen in higher dimensions. For example the large radius limit of the AdS$_4$ asymptotic symmetry algebra does not give rise to the Bondi--Metzner--Sachs BMS$_4$ algebra (see \cite{Bondi:1962px,Sachs:1962wk,Sachs:1962zza,Barnich:2009se} for the BMS algebra in 4 space-time dimensions where it was found for the first time). Hence we need another approach. This is the goal of this work. We would like to understand things directly from a flat space holographic point of view so as to have an intrinsic description of 3-dimensional flat space holography that furthermore sets the stage for 4 bulk dimensions. Nevertheless we have learned a great deal of interesting physics from the flat limit program \cite{Bagchi:2009my,Bagchi:2010zz,Barnich:2010eb,Barnich:2012aw,Krishnan:2013wta} and it allows one to compare and test results obtained using different methods.

We start our discussion of flat space holography in section \ref{sec:Sources} with a covariant description of the boundary at future null infinity $\mathcal{I}^+$. Similar statements apply at past null infinity, but we will here only consider $\mathcal{I}^+$. By employing a vielbein formalism we are able to show that the vielbein sources (the leading terms in their near boundary expansion) describe Carrollian geometry as described in \cite{Hartong:2015xda}. Next we solve the bulk equations of motion, $R_{MN}=0$, near $\mathcal{I}^+$ for general sources in section \ref{sec:eoms}. This allows us to study variations in the sense of a Schwinger functional in the form of the on-shell action with respect to the Carrollian metric-like quantities. This leads to the flat space analogue of the AdS$_3$ Brown--York boundary stress tensor that we will simply refer to as the boundary energy-momentum tensor which will be the subject of section \ref{sec:variations} where it is also shown that we have a well-posed variational problem at $\mathcal{I}^+$ without adding a Gibbons--Hawking boundary term. Finally in the last section \ref{sec:cstsources} we study what happens for constant sources at $\mathcal{I}^+$. We derive the asymptotic symmetry group by looking at the Ward identities for the boundary energy-momentum tensor and its associated conserved currents. This leads to a notion of a (generalized) conformal Killing vector for Carrollian geometry. 

One of the challenges to perform holographic calculations is that in BMS gauge, which will be assumed throughout this work, the normal to $\mathcal{I}^+$ is not manifest. This is in strong contrast with the Fefferman--Graham coordinates for asymptotically AdS space-times where the normal to boundary is parameterized by the holographic radial coordinate. Despite this drawback, BMS gauge is a very useful coordinate system for flat space holography and we will develop a general strategy to deal with the fact that the normal is not manifest and to uncover what the normal vector is. This will be discussed in sections \ref{sec:Sources} and \ref{sec:cstsources}. 

There are two types of Ward identities for the boundary energy-momentum tensor: one for boundary diffeomorphism invariance and one due to an additional local bulk diffeomorphism that acts on the Carrollian sources by a local shift. The associated Ward identity for the latter local symmetry states that the energy flux of the boundary theory must vanish and it will be shown that it is this feature which is responsible for the symmetry enhancement to the full BMS$_3$ algebra \cite{Barnich:2006av}. We also comment that there is no Ward identity for local Weyl rescalings because as we will see the boundary theory only has a one-dimensional Weyl symmetry and not a full 1+1 dimensional one. A similar restricted Weyl symmetry was observed in the context of Schr\"odinger holography in \cite{Hartong:2010ec}. We expect that a covariant description of the boundary geometry and properties of the boundary energy-momentum tensor will greatly aid in uncovering the properties of theories with BMS$_3$ symmetries \cite{Bagchi:2009ca,Bagchi:2009pe,Bagchi:2010zz,Bagchi:2012cy,Barnich:2014kra}.

\section{Sources and Carrollian geometry}\label{sec:Sources}

We will see in the coming two subsections that the geometry at (future) null infinity is described by what is called Carrollian geometry \cite{Duval:2014uoa,Duval:2014uva,Bekaert:2015xua,Hartong:2015xda}. The idea will be to work in BMS gauge and add boundary sources to it. These will be the objects that describe a general 1+1 dimensional Carrollian geometry to which the boundary field theory is coupled. In appendix \ref{app:Carrollgeometry} it will be shown that in 1+1 dimensions this geometry is in general flat (vanishing Riemann tensor) but that it has torsion. In section \ref{sec:variations} we will use variations of the on-shell action with respect to the Carrollian geometric objects to define a boundary energy-momentum tensor. One of the main challenges in realizing these goals is the fact that in BMS gauge the normal to the boundary at future null infinity is not manifest. We will see that this is closely related to the way in which local dilatations are realized in BMS gauge. The normal vector will be discussed in section \ref{subsec:normal}. In the last subsection \ref{subsec:PBHtrafos} we discuss the BMS gauge preserving diffeomorphisms.

\subsection{Adding sources to BMS gauge}

In the notation of \cite{Barnich:2006av,Bagchi:2012yk} the BMS gauge reads:
\begin{eqnarray}
g_{rr} & = & r^{-2}h_{rr}+\mathcal{O}(r^{-3})\,,\label{eq:BMSgauge1}\\
g_{ru} & = & -1+r^{-1}h_{ru}+\mathcal{O}(r^{-2})\,,\label{eq:BMSgauge2}\\
g_{r\varphi} & = & h_1(\varphi)+r^{-1}h_{r\varphi}+\mathcal{O}(r^{-2})\,,\label{eq:BMSgauge3}\\
g_{uu} & = & h_{uu}+\mathcal{O}(r^{-1})\,,\label{eq:BMSgauge4}\\
g_{u\varphi} & = & h_{u\varphi}+\mathcal{O}(r^{-1})\,,\label{eq:BMSgauge5}\\
g_{\varphi\varphi} & = & r^2+rh_{\varphi\varphi}+\mathcal{O}(1)\,,\label{eq:BMSgauge6}
\end{eqnarray}
where near boundary means large $r$. The boundary coordinates are $u$ (retarded time) and $\varphi$ (a circle coordinate with period $2\pi$). In \cite{Barnich:2006av} the condition $\partial_u h_{\varphi\varphi}=0$ and in \cite{Bagchi:2012yk} the condition $\partial_u^2h_{\varphi\varphi}=0$ is imposed. Here we will not impose these conditions. The inverse metric can be expanded as
\begin{eqnarray}
g^{rr} & = & -h_{uu}+\mathcal{O}(r^{-1})\,,\label{eq:ginvrrflat}\\
g^{ru} & = & -1-r^{-1}h_{ru}+\mathcal{O}(r^{-2})\,,\label{eq:ginvruflat}\\
g^{r\varphi} & = & r^{-2}\left(h_{u\varphi}+h_1 h_{uu}\right)+\mathcal{O}(r^{-3})\,,\\
g^{uu} & = & r^{-2}\left(h_1^2-h_{rr}\right)+\mathcal{O}(r^{-3})\,,\\
g^{u\varphi} & = & r^{-2}h_1+\mathcal{O}(r^{-3})\,,\\
g^{\varphi\varphi} & = & r^{-2}-r^{-3}h_{\varphi\varphi}+\mathcal{O}(r^{-4})\,.\label{eq:ginvphiphiflat}
\end{eqnarray}
The functions $h_{rr}$, $h_{ru}$, $h_{r\varphi}$, $h_{uu}$, $h_{u\varphi}$ and $h_{\varphi\varphi}$ depend on both $u$ and $\varphi$ and are subject to differential equations that follow from solving the bulk equations of motion and that read (see section \ref{sec:eoms} for more details)
\begin{eqnarray}
0 & = & h_{ru}+\frac{1}{2}\partial_u h_{rr}\,,\\
0 & = & \partial_u\left(h_{uu}+\partial_u h_{\varphi\varphi}\right)\,,\\
0 & = & \partial_{\varphi}h_{uu}+\partial_u\partial_\varphi h_{ru}-\partial_u^2 h_{r\varphi}+h_1\partial_u^2 h_{\varphi\varphi}-2\partial_u h_{u\varphi}\,.
\end{eqnarray}
The first equation follows from solving $R_{rr}=0$ at order $r^{-3}$ while the latter two are obtained by solving $R_{uu}=0$, $R_{u\varphi}=0$, respectively, at order $r^{-1}$. 

In \cite{Bagchi:2012yk} the choice $h_{\varphi\varphi}=h_2(\varphi)+uh_3(\varphi)$ was made implying that $h_{uu}=h_{uu}(\varphi)$. One can always achieve this by gauge fixing as we will see later. The significance of this choice in relation to the properties of the boundary at future null infinity will be discussed later in section \ref{subsec:matchingeqs}.

From now on we change notation for the coefficients indicating the order $n$ by a subscript $(n)$. We generalize the above expansions by allowing arbitrary boundary sources as follows
\begin{eqnarray}
g_{rr} & = & 2\bar\Phi r^{-2}+\mathcal{O}(r^{-3})\,,\label{eq:BMSsources1}\\
g_{r\mu} & = & -\hat\tau_\mu+r^{-1}h_{(1)r\mu}+\mathcal{O}(r^{-2})\,,\\
g_{\mu\nu} & = & r^2 h_{\mu\nu}+rh_{(1)\mu\nu}+\mathcal{O}(1)\,.\label{eq:BMSsources3}
\end{eqnarray}
We will view $\hat\tau_\mu$ and $h_{\mu\nu}$ as the sources for the energy-momentum tensor and as describing the boundary geometry. The function $\bar\Phi$ will turn out to be a pure gauge object in that we can always fix some of the bulk diffeomorphisms to set it to zero. We keep it here because it forms a natural part of the Carrollian boundary geometry\footnote{The reason it is pure gauge here, which is generally not what happens in Carrollian geometry, is due to the fact that we have an additional local symmetry whose parameter is $\chi_{(1)}^\mu$ (see appendix \ref{subsec:asymptdiffs}) that is not present in generic Carrollian geometries. This symmetry is discussed in section \ref{subsec:PBHtrafos}. We are thus dealing with a special version of Carrollian geometry and not the most generic one.}. The inverse metric is expanded as
\begin{eqnarray}
g^{rr} & = & rH_{(1)}^{rr}+\mathcal{O}(1)\,,\\
g^{r\mu} & = & v^\mu+r^{-1}H_{(1)}^{r\mu}+\mathcal{O}(r^{-2})\,,\\
g^{\mu\nu} & = & r^{-2}\bar h^{\mu\nu}+r^{-3}H_{(1)}^{\mu\nu}+\mathcal{O}(r^{-4})\,,
\end{eqnarray}
where we have
\begin{equation}
\hat\tau_\mu \bar h^{\mu\nu}=2\bar\Phi v^\nu\,,\qquad \hat\tau_\mu v^\mu=-1\,,\qquad v^\mu h_{\mu\nu}=0\,,\qquad \bar h^{\mu\rho}h_{\rho\nu}=\delta^\mu_\nu+\hat\tau_\nu v^\mu\,.
\end{equation}
Note that for general boundary sources $g^{rr}$ starts at order $r$ and not at order $1$ as in \eqref{eq:ginvrrflat}.

The fall-off conditions in \eqref{eq:BMSsources1}--\eqref{eq:BMSsources3} have been determined such that the source $\hat\tau_\mu$ can be matched onto the coefficients $\hat\tau_u=-1$ and $\hat\tau_\varphi=h_1(\varphi)$ in \eqref{eq:BMSgauge2} and \eqref{eq:BMSgauge3}, respectively and, likewise, such that the source $h_{\mu\nu}$ can reproduce the expansion of equations \eqref{eq:BMSgauge4}--\eqref{eq:BMSgauge6}. Comparing equation \eqref{eq:BMSsources3} with equations \eqref{eq:BMSgauge4}--\eqref{eq:BMSgauge6} we see that $h_{\varphi\varphi}=1$ while $h_{u\varphi}=h_{uu}=0$. Further $2\bar\Phi$ in \eqref{eq:BMSsources1} corresponds to $h_{rr}$ \eqref{eq:BMSgauge1}. The crucial difference between \eqref{eq:BMSgauge1}--\eqref{eq:BMSgauge6} and \eqref{eq:BMSsources1}--\eqref{eq:BMSsources3} is that in the latter case all the sources are arbitrary functions of the boundary coordinates $u$ and $\varphi$. We note that since the sources are more general in \eqref{eq:BMSsources1}--\eqref{eq:BMSsources3} terms that will be subleading in \eqref{eq:BMSsources1}--\eqref{eq:BMSsources3} can become  leading for special values of the sources like in \eqref{eq:BMSgauge1}--\eqref{eq:BMSgauge6}. If we look at the expansion of the inverse metric we likewise see that for special values of the sources $v^\mu$ and $\bar h^{\mu\nu}$ we reproduce \eqref{eq:ginvruflat}--\eqref{eq:ginvphiphiflat}. The difference being again that the sources $v^\mu$ and $\bar h^{\mu\nu}$ are now allowed to be arbitrary functions of the boundary coordinates. This has a consequence for the expansion of the inverse metric component $g^{rr}$ which now starts at order $r$ (something that follows from the equations of motion at leading order in $r$). The coefficient is denoted by $H_{(1)}^{rr}$ where the subscript $(1)$ indicates that it is first order in boundary derivatives.

\subsection{From bulk to boundary vielbeins: Carrollian geometry}\label{subsec:bdrygeometry}

In order to understand what kind of geometry the sources $\hat\tau_\mu$, $h_{\mu\nu}$ and $\bar\Phi$ describe we need to understand their properties. To this end it will prove very convenient to write the metric in BMS gauge in vielbein formalism. This allows us to work out the tangent space properties of the boundary geometry. With that information at hand we can then determine the boundary geometry from a procedure known as gauging space-time symmetries like it was done for the case of Lifshitz space-times in \cite{Christensen:2013lma,Christensen:2013rfa,Hartong:2014oma,Bergshoeff:2014uea,Hartong:2015wxa}. We will first derive the tangent space transformations of the boundary vielbeins. After this the reader is referred to appendix \ref{app:Carrollgeometry} for the details of the gauging procedure.

We will employ a null-bein basis of two null vectors $U$ and $V$ that are normalized such that $g^{MN}U_MV_N=-1$, where one of them, $U_M$, is taken to be the normal vector to the boundary at future null infinity\footnote{Bulk space-time coordinates are indicated by $x^M$ and boundary space-time coordinates at $\mathcal{I}^+$  by $x^\mu$.} $\mathcal{I}^+$. The third remaining vielbein $E$ is a unit spacelike vector that is orthogonal to $U$ and $V$. We thus decompose the metric as
\begin{equation}\label{eq:vielbeins}
ds^2=-2UV+EE\,.
\end{equation}
At this stage we do not know the explicit form of the normal vector to $\mathcal{I}^+$ so we will assume it is an arbitrary hypersurface orthogonal (HSO) null vector. Further below we will confirm that one can make this assumption without loss of generality (WLOG). This is important because this guarantees that among the vielbein decompositions of the bulk metric we will find the one relevant for the vielbein description of the geometry close to $\mathcal{I}^+$. It is then a matter of extracting that information. This will be done in section \ref{sec:variations} by demanding that there is a well-posed variational problem.

The decomposition \eqref{eq:vielbeins} has a local $SO(1,2)$ symmetry acting on $U$, $V$ and $E$. One of these is the boost transformation 
\begin{eqnarray}
U' & = & \bar\Lambda U\,,\label{eq:LLboost1}\\
V' & = & \bar\Lambda^{-1}V\,.\label{eq:LLboost2}
\end{eqnarray}
Since $U_M$ is a HSO null vector we can rescale it with any function and it will still be a HSO null vector. That is to say that we can use the $\bar\Lambda$ freedom to fix one of the components of $U_M$. We will assume that $U_r\neq 0$ so we can always set $U_r=1$. Rather than doing that to all orders, we just impose this condition at leading order. Later we will set $U_r=1$ to all orders, but it allows us to see certain properties related to transformations under bulk diffeomorphisms more clearly (see the last paragraph of section \ref{subsec:asymptdiffs}). 

We are now in a position to work out the vielbein expansion for large $r$ which is given by
\begin{eqnarray}
U_r & = & 1+\mathcal{O}(r^{-1})\,,\label{eq:falloff1}\\
U_\mu & = & rU_{(1)\mu}+\mathcal{O}(1)\,,\\
V_r & = & r^{-2}\tau_\mu M^\mu+\mathcal{O}(r^{-3})\,,\\
V_\mu & = & \tau_\mu+\mathcal{O}(r^{-1})\,,\\
E_r & = & r^{-1}e_\nu M^\nu+\mathcal{O}(r^{-2})\,,\\
E_\mu & = & re_\mu+\mathcal{O}(1)\,,\label{eq:falloff6}
\end{eqnarray}
where $U_{(1)\mu}$, $\tau_\mu$, $M^\mu$ and $e_\mu$ are independent of the coordinate $r$. The inverse vielbeins are expanded as
\begin{eqnarray}
U^r & = & \mathcal{O}(r)\,,\\
U^\mu & = & v^\mu+\mathcal{O}(r^{-1})\,,\\
V^r & = & -1+\mathcal{O}(r^{-1})\,,\\
V^\mu & = & r^{-2}M^\mu+\mathcal{O}(r^{-3})\,,\\
E^r & = & \mathcal{O}(1)\,,\\
E^\mu & = & r^{-1}e^\mu+\mathcal{O}(r^{-2})\,,
\end{eqnarray}
where we have
\begin{equation}
v^\mu\tau_\mu=-1\,,\quad v^\mu e_\mu =0\,,\quad e^\mu\tau_\mu=0\,,\quad e^\mu e_\mu=1\,.
\end{equation}

We now work out under which local transformations the boundary vielbeins $\tau_\mu$ and $e_\mu$ as well as the vector $M^\mu$, that are defined via the above fall-off conditions, transform. 
Since $U_M$ is the normal vector to the boundary the bulk local Lorentz transformations that keep $U_M$ fixed are going to give rise to the local tangent transformations of the boundary vielbeins. The idea here is entirely analogous to the way torsional Newton--Cartan geometry was shown to be the geometric description of the boundary of asymptotically locally Lifshitz space-times in \cite{Christensen:2013lma,Christensen:2013rfa,Hartong:2014oma,Bergshoeff:2014uea,Hartong:2015wxa}. For this purpose it is convenient to temporarily label $U$, $V$ and $E$ as
\begin{equation}
E^+=U\,,\qquad E^-=V\,,\qquad E^2=E\,,
\end{equation}
so that we have
\begin{equation}
ds^2=\eta_{AB}E^AE^B=-2E^+E^-+E^2 E^2\,.
\end{equation}
Consider the local Lorentz transformation $E'^A=\Lambda^A{}_B E^B$ where $\eta_{AB}=\Lambda^C{}_A\Lambda^D{}_B\eta_{CD}$. Imposing that $U'=E'^+=E^+=U$ we find $\Lambda^+{}_+=1$, $\Lambda^+{}_-=0$, $\Lambda^+{}_2=0$. The invariance condition $\eta_{AB}=\Lambda^C{}_A\Lambda^D{}_B\eta_{CD}$ then tells us that $\Lambda^2{}_2=1$, $\Lambda^2{}_-=0$, $\Lambda^-{}_-=1$ and $\Lambda^-{}_2=\Lambda^2{}_+$. Hence there is one free function $\Lambda^2{}_+\equiv\Lambda$ corresponding to the transformation (null rotation)
\begin{eqnarray}
U' & = & U\,,\label{eq:nullrot1}\\
V' & = & V+\Lambda E+\frac{1}{2}\Lambda^2 U\,,\label{eq:nullrot2}\\
E' & = & E+\Lambda U\,.\label{eq:nullrot3}
\end{eqnarray}
We note that using the fact that we can set $U_r=1$, it is always possible to set $E_r=0$ by fixing null rotations.

In order to respect the fall-off conditions for the vielbeins \eqref{eq:falloff1}--\eqref{eq:falloff6} after acting with the local Lorentz transformation that keeps $U_M$ fixed, i.e. the transformed vielbeins should have the same near boundary expansion as the untransformed ones except that the boundary vielbeins $\tau_\mu$, $e_\mu$ as well as $M^\mu$ may have transformed, we need that the parameter $\Lambda$ falls off like
\begin{equation}
\Lambda=r^{-1}\lambda+\mathcal{O}(r^{-2})\,.
\end{equation}
In other words we require that
\begin{equation}
E'_\mu=re'_\mu+\ldots=E_\mu+\Lambda U_\mu\,,
\end{equation}
where $E$ and $U$ are expanded as in \eqref{eq:falloff1}--\eqref{eq:falloff6} and likewise for the other vielbeins. The null rotation \eqref{eq:nullrot1}--\eqref{eq:nullrot3} acts on the sources as
\begin{eqnarray}
e'_\mu & = & e_\mu\,,\\
\tau'_\mu & = & \tau_\mu+\lambda e_\mu\,,\\
e'_\mu M'^\mu & = & e_\mu M^\mu+\lambda\,,\\
\tau'_\mu M'^\mu & = & \tau_\mu M^\mu +\lambda e_\mu M^\mu+\frac{1}{2}\lambda^2\,,
\end{eqnarray}
from which we conclude that 
\begin{eqnarray}
M'^\mu & = & M^\mu+\lambda e^\mu+\frac{1}{2}\lambda^2 v^\mu\,,\label{eq:Carrtrafo1}\\
v'^\mu & = & v^\mu\,,\\
e'^\mu & = & e^\mu+\lambda v^\mu \,.\label{eq:Carrtrafo3}
\end{eqnarray}
We thus see that $\tau_\mu$, $e_\mu$ and $M^\mu$ transform like in Carrollian geometry as reviewed in appendix \ref{app:Carrollgeometry}. In section \ref{subsec:PBHtrafos} we will see that $\tau_\mu$, $e_\mu$ and $M^\mu$ also transform under boundary diffeomorphisms as well as under two additional local transformations: local dilatations with parameter $\Lambda_D$ and local shift transformations with parameter $\chi_{(1)}^\mu$. The local dilatations can be incorporated by looking at the Lifshitz--Carroll algebra \cite{Gibbons:2009me,Bergshoeff:2015wma}. It would be interesting to find out which group containing Lifshitz--Carroll as a subgroup is responsible for these extra local shift transformations. 

Using the transformations given above for the vielbeins and their inverses we can use $M^\mu$ to build the following set of Carroll boost invariant objects
\begin{eqnarray}
\hat\tau_\mu & = & \tau_\mu-h_{\mu\nu}M^\nu\,,\\
\bar h^{\mu\nu} & = & h^{\mu\nu}-v^\mu M^\nu-v^\nu M^\mu\,,\\
\bar\Phi & = & -\tau_\mu M^\mu+\frac{1}{2}h_{\mu\nu}M^\nu M^\nu\,,\\
\hat e^\mu & = & e^\mu-v^\mu e_\nu M^\nu\,,\\
\hat h^{\mu\nu} & = & \hat e^\mu\hat e^\nu=\bar h^{\mu\nu}+2\bar\Phi v^\mu v^\nu\,,
\end{eqnarray}
where we defined
\begin{equation}
h_{\mu\nu}=e_\mu e_\nu\,,\qquad h^{\mu\nu}=e^\mu e^\nu\,.
\end{equation}
Comparing equations \eqref{eq:BMSsources1}--\eqref{eq:BMSsources3} with the BMS gauge of the previous subsection, eqs. \eqref{eq:BMSgauge1}--\eqref{eq:BMSgauge6}, we conclude that the latter can now be understood as corresponding to the following choice of boundary Carrollian sources
\begin{eqnarray}
\hat\tau_\mu & = & \delta_\mu^u-h_1\delta_\mu^\varphi\,,\\
e_\mu & = & \delta_\mu^\varphi\,,\\
\hat e^\mu & = & h_1\delta^\mu_u+\delta^\mu_\varphi\,,\\
v^\mu & = & -\delta^\mu_u\,,\\
\bar\Phi & = & \frac{1}{2}h_{rr}\,,\\
\bar h^{uu} & = & h_1^2-h_{rr}\,,\qquad\bar h^{u\varphi}=h_1\,,\qquad\bar h^{\varphi\varphi}=1\,.
\end{eqnarray}

\subsection{Boundary normal vector}\label{subsec:normal}

In BMS gauge the location of the boundary at future null infinity is not manifest, so it is not straightforward to write down the normal vector $U_M$. Such a situation does not arise in AdS holography in Fefferman--Graham gauge. The way we will deal with this is that we impose on $U_M$ the properties that it is a null hypersurface orthogonal vector which it must be in any case. The 1-form $U_M$ is then normal to a very general class of hypersurfaces that must contain future null infinity but that will also contain other hypersurfaces. One of the challenges will be to derive a set of conditions that identifies which vector $U_M$ we should take. Later we will derive a set of equations called the matching equations that will determine $U_M$ from given functions in the BMS gauge.

The large $r$ expansion of $U_M$ is given by
\begin{eqnarray}
U_r & = & 1+r^{-1}U_{(1)r}+\mathcal{O}(r^{-2})\,,\label{eq:expUr}\\
U_\mu & = & rU_{(1)\mu}+U_{(2)\mu}+\mathcal{O}(r^{-1})\,.\label{eq:expUmu}
\end{eqnarray}
This follows from the results of appendix \ref{subsec:vielbeins}. Since $U_M$ is HSO it satisfies the Frobenius integrability condition
\begin{equation}
U_M\left(\partial_N U_P-\partial_P U_N\right)+U_P\left(\partial_M U_N-\partial_N U_M\right)+U_N\left(\partial_P U_M-\partial_M U_P\right)=0\,,
\end{equation}
which using \eqref{eq:expUr} and \eqref{eq:expUmu} becomes asymptotically
\begin{eqnarray}
0 & = & \partial_\mu U_{(1)\nu}-\partial_\nu U_{(1)\mu}\,,\label{eq:intbU1}\\
0 & = & \partial_\mu U_{(2)\nu}-\partial_\nu U_{(2)\mu}+U_{(1)\mu}\left(U_{(2)\nu}+\partial_\nu U_{(1)r}\right)-U_{(1)\nu}\left(U_{(2)\mu}+\partial_\mu U_{(1)r}\right)\,.\label{eq:intbU2}
\end{eqnarray}
These can be solved in terms of two new scalar functions $U_{(1)}$ and $U_{(2)}$ such that
\begin{eqnarray}
U_{(1)\mu} & = & \partial_\mu U_{(1)}\,,\label{eq:U1}\\
U_{(2)\mu} & = & -\partial_\mu U_{(1)r}+e^{-U_{(1)}}\partial_\mu U_{(2)}\,.\label{eq:U2}
\end{eqnarray}

From the vielbein expansions of section \ref{subsec:vielbeins} we learn that the normal $U_M$ must obey the following conditions (due to it being a null vector)
\begin{eqnarray}
\hspace{-.5cm}v^\mu U_{(1)\mu} & = & -K\,,\label{eq:conditionU1}\\
\hspace{-.5cm}v^\mu U_{(2)\mu}+KU_{(1)r}+\frac{1}{2}\left(\hat e^\mu U_{(1)\mu}-\hat e^\mu\mathcal{L}_v\hat\tau_\mu\right)^2 & = & \frac{1}{2}v^\mu v^\nu h_{(2)\mu\nu}-\bar\Phi K^2-Kv^\mu\partial_\mu\bar\Phi\,,\label{eq:conditionU2}
\end{eqnarray}
where $K$ is the trace of the boundary extrinsic curvature, i.e. $K=h^{\mu\nu}K_{\mu\nu}$ with $K_{\mu\nu}=-\frac{1}{2}\mathcal{L}_v h_{\mu\nu}$ in which $\mathcal{L}_v$ denotes the Lie derivative along $v^\mu$. We can always find $U_{(1)}$ and $U_{(2)}$ such that these equations are solved. We thus conclude that we can always take $U_M$ to be null and HSO.

Given that in the vielbein decomposition we can WLOG take $U_M$ to be HSO it becomes relevant to ask to which hypersurfaces $U_M$ is the normal vector. Solving the Frobenius integrability condition we can always write $U_M=N\partial_M F$. The asymptotic expansions of $N$ and $F$ are given by
\begin{eqnarray}
N & = & e^{-U_{(1)}}\left(1+r^{-1}U_{(1)r}+\mathcal{O}(r^{-2})\right)\,,\\
F & = & r e^{U_{(1)}}+U_{(2)}-e^{U_{(1)}}U_{(1)r}+\mathcal{O}(r^{-1})\,.
\end{eqnarray}
The hypersurface to which $U_M$ is normal is therefore given by 
\begin{equation}\label{eq:hypersurfaces}
F=r e^{U_{(1)}}+U_{(2)}-e^{U_{(1)}}U_{(1)r}+\mathcal{O}(r^{-1})=\text{cst}\,. 
\end{equation}
The sources $\hat\tau_\mu$, $e_\mu$ and $\bar\Phi$ are defined on the null hypersurface $F=\text{cst}$ at large $r$.

\subsection{Diffeomorphism redundancies of BMS gauge}\label{subsec:PBHtrafos}

For large $r$ (near boundary) the metric and normal vector are expanded as in \eqref{eq:BMSsources1}--\eqref{eq:BMSsources3} and \eqref{eq:expUr}, \eqref{eq:expUmu}, respectively. There is a very large amount of gauge redundancy in these expansions which we will now show. This allows us to identify useful gauge choices as well as additional local transformations of our sources (already alluded to above). Finally for later purposes it will also allow us to understand the expansion of subleading components of the metric that are related to the boundary energy-momentum tensor, but this will not be discussed in this subsection. 

In appendix \ref{subsec:asymptdiffs} we work out the general BMS gauge preserving diffeomorphisms in the presence of sources. Here we highlight some of the most important results. The sources transform under bulk diffeomorphisms generated by
\begin{eqnarray}
\xi^r & = & r\Lambda_D+\xi^r_{(1)}+r^{-1}\xi^r_{(2)}+\mathcal{O}(r^{-2})\,,\\
\xi^\mu & = & \chi^\mu+r^{-1}\chi^\mu_{(1)}+r^{-2}\chi^\mu_{(2)}+\mathcal{O}(r^{-3})\,,
\end{eqnarray}
as
\begin{eqnarray}
\delta\hat\tau_\mu & = & \Lambda_D\hat\tau_\mu+\mathcal{L}_\chi\hat\tau_\mu+h_{\mu\rho}\chi_{(1)}^\rho\,,\label{eq:localtrafosources1}\\
\delta e_\mu & = & \Lambda_D e_\mu+\mathcal{L}_\chi e_\mu\,,\label{eq:localtrafosources2}\\
\delta M^\mu & = & -\Lambda_D M^\mu+\mathcal{L}_\chi M^\mu-\chi_{(1)}^\mu\,.
\end{eqnarray}
We see that the local dilatations have a dynamical exponent $z=1$. This is the reason why the near boundary Taylor expansion in $r$ is also a derivative expansion. Every order in $r$ further down the expansion means adding one derivative. We will later in section \ref{sec:variations} see that the role of local dilatations is very different from the AdS case where they give rise to local Weyl invariance of the boundary theory (up to an anomaly). Later we will find that there is no analogue of Weyl invariance of the boundary theory at $\mathcal{I}^+$. The $\chi_{(1)}^\mu$ transformations play an important role in the realization of the BMS symmetries as those BMS gauge preserving diffeomorphisms that leave the Carrollian sources invariant (see section \ref{subsec:BMS}). The $\chi_{(1)}^\mu$ transformations allow us to remove $M^\mu$ entirely by gauge fixing. The freedom to set $M^\mu=0$ is the geometric counterpart of Carrollian boost invariance of some action defined on a Carrollian geometry. Here one can think of the analogy with Lorentzian geometry where the freedom to perform local Lorentz transformations on the vielbeins is the geometric counterpart of Lorentz invariance of some field theory defined on a Lorentzian background.

Of the subleading terms at first order we give here only the transformations of $\hat e^\mu h_{(1)r\mu}$ and $\hat h^{\mu\nu}h_{(1)\mu\nu}$ because as we will see in section \ref{subsec:LO+NLO} the remaining components of $h_{(1)r\mu}$ and $h_{(1)\mu\nu}$ are fixed by solving the bulk equations of motion in terms of derivatives of the sources. The transformations of $\hat e^\mu h_{(1)r\mu}$ and $\hat h^{\mu\nu}h_{(1)\mu\nu}$ are given by
\begin{eqnarray}
\delta\left(\hat e^\mu h_{(1)r\mu}\right) & = & -\Lambda_D\hat e^\mu h_{(1)r\mu}+\mathcal{L}_\chi\left(\hat e^\mu h_{(1)r\mu}\right)+2\bar\Phi\hat e^\mu\partial_\mu\Lambda_D-\hat e^\mu\mathcal{L}_{\chi_{(1)}}\hat\tau_\mu\nonumber\\
&&-\hat e^\mu h_{(1)\mu\nu}\chi_{(1)}^\nu+v^\mu h_{(1)r\mu}e_\nu\chi_{(1)}^\nu-2e_\mu\chi_{(2)}^\mu\,,\\
\delta\left(\hat h^{\mu\nu}h_{(1)\mu\nu}\right) & = & -\Lambda_D \hat h^{\mu\nu}h_{(1)\mu\nu}+\mathcal{L}_\chi\left(\hat h^{\mu\nu}h_{(1)\mu\nu}\right)+\hat h^{\mu\nu}\mathcal{L}_{\chi_{(1)}}h_{\mu\nu}\nonumber\\
&&+2v^\mu\hat e^\nu h_{(1)\mu\nu}e_\rho X^\rho_{(1)}+2\xi^r_{(1)}\,.
\end{eqnarray}
In order to obtain the transformation of the normal vector we use \eqref{eq:U1} and \eqref{eq:U2} as well as \eqref{eq:trafoU1r}--\eqref{eq:trafoU2} leading to
\begin{eqnarray}
\delta U_{(1)r} & = & -\Lambda_D U_{(1)r}+ \mathcal{L}_\chi U_{(1)r}-U_{(1)\mu}\chi_{(1)}^\mu+\alpha_{(1)}\,,\\
\delta U_{(1)} & = & \Lambda_D+\mathcal{L}_\chi U_{(1)}\,,\\
\delta U_{(2)} & = & \mathcal{L}_\chi U_{(2)}+e^{U_{(1)}}\left(\xi^r_{(1)}+\alpha_{(1)}\right)\,.
\end{eqnarray}
The parameter $\alpha_{(1)}$ is included due to the fact that we perform a local Lorentz boost on $U_M$ together with a diffeomorphism, as explained in appendix \ref{subsec:asymptdiffs} in equation \eqref{eq:trafoUM}. We can fix $U_{(1)r}=0$ by setting $\alpha_{(1)}=U_{(1)\mu}\chi_{(1)}^\mu$. 

We can gauge fix $M^\mu$ to be zero up to local Carroll boosts. In particular we can set $\bar\Phi=0$ (see \eqref{eqLdiffeobarPhi}). This fixes $\hat\tau_\mu\chi^\mu_{(1)}$. We can also set $\hat e^\mu h_{(1)r\mu}=0$ and $h_{(1)rr}=0$, which appears subleading to $\bar\Phi$ in the expansion of $g_{rr}$ at order $r^{-3}$ and whose transformation is given in \eqref{eq:h1rr}, by fixing $\chi_{(2)}^\mu$. We could in principle also gauge fix $\hat h^{\mu\nu}h_{(1)\mu\nu}=0$ by using the $\xi^r_{(1)}$ transformation. However the $\xi^r_{(1)}$ transformation also acts on the normal vector and so we will not fix it before we understand the implications it has for the normal vector. We also note that the local dilatations can be used to fix $U_{(1)}$ to be a constant leaving us with just global scale transformations, i.e. constant $\Lambda_D$. Again we will not do this because we would like to keep the freedom to perform local rescalings as free as possible since this might tell us something useful about the dual field theory, but it is interesting to observe that we have enough diffeomorphism freedom to set both $U_{(1)}$ and $U_{(2)}$ equal to constants by fixing both $\Lambda_D$ and $\xi_{(1)}^r$ transformations. In this gauge the vector $\partial_M r$ is asymptotically null because in this gauge $g^{rr}$ goes to zero as $r\rightarrow\infty$. This follows from \eqref{eq:invgrr}, \eqref{eq;H1rr}, \eqref{eq:U1}, \eqref{eq:H2rr} and \eqref{eq:U2}.

\section{On-shell expansions near null infinity}\label{sec:eoms}

In this section we will solve the equations of motion $R_{MN}=0$ order by order for large $r$. We will do this in two steps. First we will determine the solution at next-to-leading order (NLO) and subsequently up to  N$^3$LO where we will find the on-shell differential equations involving $h_{(2)\mu\nu}$ that we will relate to Ward identities for the on-shell action at $\mathcal{I}^+$. We use the definition that N$^k$LO means solving $R_{MN}=0$ at the following orders
\begin{eqnarray}
R_{rr} & = & o(r^{-2-k})\,,\\
R_{r\mu} & = & o(r^{-k})\,,\\
R_{\mu\nu} & = & o(r^{2-k})\,.
\end{eqnarray}

The fact that we have a covariant description of the boundary allows us to test the assumption that the large $r$ expansion is a Taylor series because if at some subleading order we find that the Taylor expansion ansatz leads to a restriction on the sources we should look for a more general expansion (containing possibly $\log r$ terms) in order to keep the sources unconstrained. We will see that no $\log$ terms are necessary and that a Taylor series expansion is adequate\footnote{This is not true in higher dimensions where the Taylor expansion leads to constraints on the boundary sources.}.

\subsection{Solving the equations of motion at LO and NLO}\label{subsec:LO+NLO}

Using the results of appendix \ref{sec:expuptoNLO} we find that demanding that $R_{rr}$ vanishes at order $r^{-3}$ tells us that
\begin{equation}
v^\mu h_{(1)r\mu}=-v^\mu\partial_\mu\bar\Phi+2\bar\Phi K+2\bar\Phi v^\mu v^\nu h_{(1)\mu\nu}\,.
\end{equation}
Demanding that $R_{r\mu}$ vanishes at order $r^{-1}$ gives
\begin{equation}
v^\mu h_{(1)\mu\nu}=-2\hat\tau_\nu K-2\hat\tau_\nu v^\rho v^\sigma h_{(1)\rho\sigma}+\mathcal{L}_v\hat\tau_\nu\,.
\end{equation}
Contracting this equation with $v^\nu$ we obtain
\begin{equation}
v^\mu v^\nu h_{(1)\mu\nu}=-K\,,
\end{equation}
so that $v^\mu h_{(1)r\mu}$ and $v^\mu h_{(1)\mu\nu}$ simplify to
\begin{eqnarray}
v^\mu h_{(1)r\mu} & = & -v^\mu\partial_\mu\bar\Phi-2\bar\Phi K\,,\label{eq:rel1}\\
v^\mu h_{(1)\mu\nu} & = & 2\hat\tau_\nu K+\mathcal{L}_v\hat\tau_\nu\,.\label{eq:rel2}
\end{eqnarray}
From now on we will always assume that equations \eqref{eq:rel1} and \eqref{eq:rel2} are obeyed. The $R_{\mu\nu}$ equation at order $r$ is satisfied automatically\footnote{This is not true in higher dimensions if we assume a Taylor expansion.}.

\subsection{Solving the equations of motion at N$^2$LO and N$^3$LO}\label{subsec:N2LO}

At N$^2$LO we are demanding that $R_{rr}$ vanishes at order $r^{-4}$. With the metric expansion as given in section \ref{subsec:metricexp} we are not able to compute\footnote{If we were to expand the metric to sufficiently high orders so that we can compute $R_{rr}$ at order $r^{-4}$ we would find that it determines a certain combination of subleading coefficients.} $R_{rr}$ at order $r^{-4}$ so we will continue with demanding that $R_{r\mu}$ vanishes at order $r^{-2}$ and that $R_{r\mu}$ vanishes at order $r^0$ which we can compute with the orders given in  section \ref{subsec:metricexp}. It turns out that, using the results of section \ref{subsec:connections}, the equations of motion for $R_{r\mu}$ and $R_{r\mu}$ at N$^2$LO are satisfied automatically.

The interesting equations appear at N$^3$LO where we will find two differential equations for subleading coefficients. The expansions given in appendix \ref{app:asymptotics} are designed such that we can compute $v^\mu v^\nu R_{\mu\nu}$ and $v^\mu \hat e^\nu R_{\mu\nu}$ at order $r^{-1}$, but not $\hat e^\mu \hat e^\nu R_{\mu\nu}$ or $R_{r\mu}$. The reason for this is that the equations for $\hat e^\mu \hat e^\nu R_{\mu\nu}$ or $R_{r\mu}$ at N$^3$LO just fix subleading coefficients and do not lead to any differential equations that take the form of Ward identities that must be obeyed. 

Setting $v^\mu v^\nu R_{\mu\nu}$ to zero at order $r^{-1}$ leads to
\begin{eqnarray}
0 & = & \left(v^\rho\partial_\rho-2K\right)\left(\frac{1}{2} v^\mu v^\nu h_{(2)\mu\nu}-K^2\bar\Phi -Kv^\mu\partial_\mu\bar\Phi-\frac{1}{2}v^\mu\partial_\mu\left(\hat h^{\lambda\kappa}h_{(1)\lambda\kappa}\right)\right.\nonumber\\
&&\left.+\frac{1}{2}K\hat h^{\mu\nu}h_{(1)\mu\nu}+\hat e^\mu\partial_\mu\left(\hat e^\nu\mathcal{L}_v\hat\tau_\nu\right)+\frac{1}{2}\left(\hat e^\mu\mathcal{L}_v\hat\tau_\mu\right)^2\right)\nonumber\\
&&+e^{-1}\partial_\mu\left[e\hat h^{\mu\nu}\left(K\mathcal{L}_v\hat\tau_\nu+\partial_\nu K\right)\right]\,.\label{eq:WI1}
\end{eqnarray}
Setting $v^\mu \hat e^\nu R_{\mu\nu}$ to zero at order $r^{-1}$ leads to
\begin{eqnarray}
0 & = & \left(\hat e^\rho\partial_\rho+2\hat e^\rho\mathcal{L}_v\hat\tau_\rho\right)\left(-\frac{1}{2}v^\mu v^\nu h_{(2)\mu\nu}+K^2\bar\Phi+Kv^\mu\partial_\mu\bar\Phi+\frac{1}{2}v^\mu\partial_\mu\left(\hat h^{\lambda\kappa}h_{(1)\lambda\kappa}\right)\right.\nonumber\\
&&\left.-\frac{1}{2}K\hat h^{\lambda\kappa}h_{(1)\lambda\kappa}\right)+\left(v^\rho\partial_\rho-2K\right)\left(\frac{1}{2}\hat e^\mu\partial_\mu\left(v^\nu\partial_\nu\bar\Phi\right)+\frac{1}{2}\left(\hat e^\mu\mathcal{L}_v\hat\tau_\mu\right)v^\nu\partial_\nu\bar\Phi\right.\nonumber\\
&&\left.+\bar\Phi\hat e^\mu\partial_\mu K+\bar\Phi K\hat e^\mu\mathcal{L}_v\hat\tau_\mu+\frac{1}{2}v^\mu\partial_\mu\left(\hat e^\nu h_{(1)r\nu}\right)+\frac{1}{2}K\hat e^\mu h_{(1)r\mu}\right.\nonumber\\
&&\left.-\frac{1}{2}\hat e^\mu\partial_\mu\left(\hat h^{\lambda\kappa}h_{(1)\lambda\kappa}\right)-\frac{1}{2}\left(\hat e^\mu\mathcal{L}_v\hat\tau_\mu\right)\hat h^{\lambda\kappa}h_{(1)\lambda\kappa}+\hat e^\mu v^\nu h_{(2)\mu\nu}\right)\,.\label{eq:WI2}
\end{eqnarray}

We thus find two equations that cannot be written as an expression for a certain coefficient and that we would like to interpret as Ward identities for some local symmetry of the on-shell action. This will be the subject of the next section.

We remind the reader that a similar situation occurs when solving the bulk equations of motion for an action with a negative cosmological constant, i.e. for asymptotically AdS$_3$ solutions. If we choose Fefferman--Graham gauge the boundary energy-momentum tensor appears at NLO. This quantity is not fully determined by the equations of motion but instead has to satisfy certain Ward identities (see e.g. \cite{Balasubramanian:1999re,deHaro:2000vlm}). Here we will likewise interpret \eqref{eq:WI1} and \eqref{eq:WI2} as the Ward identities for a boundary energy-momentum tensor.

\section{Well-posed variational principle}\label{sec:variations}

The goal of this section is to set up a well-posed variational problem for variations that vanish at $\mathcal{I}^+$. A similar problem was studied using spatial slices in \cite{Detournay:2014fva} where they work in a radial gauge with a unit spacelike normal vector (and Wick rotated geometries). We will find that the variational problem at $\mathcal{I}^+$ does not require any boundary terms\footnote{This is different in \cite{Detournay:2014fva} where the use of spatial cut-off hypersurfaces requires the use of a Gibbons--Hawking boundary term.}. The next step is to define a boundary energy-momentum tensor and to derive its Ward identities. We start by defining a boundary integration measure at $\mathcal{I}^+$.

\subsection{Boundary integration measure}\label{subsec:intmeasure}

Consider the 3D bulk Levi--Civit\`a tensor written in terms of the bulk vielbeins as
\begin{equation}
\epsilon_{MNP}=\left(V_M E_N-V_N E_M\right)U_P+\left(E_M U_N-E_N U_M\right)V_P+\left(U_M V_N-U_N V_M\right)E_P\,.
\end{equation}
The boundary integration measure is given by
\begin{equation}
\frac{1}{2}\epsilon_{MNP} dx^M\wedge dx^N V^P\vert_{\partial{\mathcal{M}}}\,,
\end{equation}
where $\partial{\mathcal{M}}$ indicates the hypersurface $F=\text{cst}$ to which $U_M$ is orthogonal, equation \eqref{eq:hypersurfaces}. This is invariant under the null rotations \eqref{eq:nullrot1}--\eqref{eq:nullrot3} using that $U_Mdx^M$ vanishes on the boundary. Expanding $\frac{1}{2}\epsilon_{MNP}V^P dx^M\wedge dx^N$ for large $r$ we find
\begin{eqnarray}
\frac{1}{2}\epsilon_{MNP}V^P dx^M\wedge dx^N & = & \frac{1}{2}\epsilon_{\mu\nu}dx^\mu\wedge dx^\nu\left(r+\frac{1}{2}\hat h^{\rho\sigma}h_{(1)\rho\sigma}-v^\rho\partial_\rho\bar\Phi-U_{(1)r}\right)\nonumber\\
&&+\epsilon_{\mu\nu}M^\nu dx^\mu\wedge\left(\frac{dr}{r}+U_{(1)\rho}dx^\rho\right)\nonumber\\
&&+\mathcal{O}(r^{-1})dx^\mu\wedge dx^\nu+\mathcal{O}(r^{-2})dx^\mu\wedge dr\,.
\end{eqnarray}
When evaluating this on the surface $F=\text{cst}$ the second line vanishes up to the order that we are expanding. Hence we have
\begin{equation}\label{eq:expintmeasure}
\frac{1}{2}\epsilon_{MNP}V^P dx^M\wedge dx^N\vert_{\partial{\mathcal{M}}}=\frac{1}{2}\epsilon_{\mu\nu}dx^\mu\wedge dx^\nu\left(r+\frac{1}{2}\hat h^{\rho\sigma}h_{(1)\rho\sigma}-v^\rho\partial_\rho\bar\Phi-U_{(1)r}+\mathcal{O}(r^{-1})\right)\,.
\end{equation}
This result will be important further below.

\subsection{Dirichlet conditions without Gibbons--Hawking boundary term}\label{subsec:GHbdryterm}

The bulk action is 
\begin{equation}
S=\int d^3x\sqrt{-g}R\,.
\end{equation}
Its variation is given by
\begin{equation}
\delta S=\int d^3x\sqrt{-g}G_{MN}\delta g^{MN}+\int d^3x\partial_M\left(\sqrt{-g}J^M\right)\,,
\end{equation}
where the current $J^M$ is given by
\begin{equation}
J^M=g^{NP}\delta\Gamma^M_{NP}-g^{MP}\delta\Gamma_{PN}^N\,.
\end{equation}
Ignoring the first term containing the bulk equation of motion we can write
\begin{equation}\label{eq:variationaction}
\delta S=\int_{\partial\mathcal{M}} \frac{1}{2}\epsilon_{MNP}dx^M\wedge dx^N J^P=-\frac{1}{2}\int_{\partial\mathcal{M}}\epsilon_{MNP}dx^M\wedge dx^N V^P U_QJ^Q\,,
\end{equation}
where we used the fact that only the component of $J^P$ along $V^P$ contributes since all the others vanish when evaluated on the boundary using $U_Mdx^M\vert_{\partial\mathcal{M}}=0$. Note that the right hand side of \eqref{eq:variationaction} is invariant under the local Lorentz boost \eqref{eq:LLboost1} and \eqref{eq:LLboost2} that rescales the normal vector. 

We have a well-posed Dirichlet variational problem if $\delta S$ vanishes on-shell for variations of the sources that vanish on the boundary $\partial\mathcal{M}$. Here of course we are interested in the part of the boundary that corresponds to $\mathcal{I}^+$. Before getting into that let us consider what the GH boundary term looks like regardless the question of whether we actually need one or not.

It would have to be of the following form 
\begin{equation}
S_{\text{GH}}=\alpha\int_{\partial\mathcal{M}}\frac{1}{2}\epsilon_{MNP}dx^M\wedge dx^N V^P K_{\text{GH}}\,,
\end{equation}
where $\alpha$ is a constant and where $K_{\text{GH}}$ is a scalar quantity involving the covariant derivative of the normal vector such that $S_{\text{GH}}$ is invariant under the local Lorentz boost \eqref{eq:LLboost1} and \eqref{eq:LLboost2} and the null rotation \eqref{eq:nullrot1}--\eqref{eq:nullrot3}. The only term that obeys these requirements is
\begin{equation}
K_{\text{GH}}=E^M E^N\nabla_M U_N\,.
\end{equation}
In order to prove the invariance we need to use that $U_M$ is null and HSO. It can be shown that on-shell\footnote{On-shell we have
\begin{equation}
K_{\text{GH}}=r^{-1}Z_{(2)}+\mathcal{O}(r^{-2})\,,
\end{equation}
where
\begin{equation}
Z_{(2)}=\left(\hat e^\mu U_{(1)\mu}\right)^2-2\hat e^\mu U_{(1)\mu}\hat e^\nu\mathcal{L}_v\hat\tau_\nu+v^\mu U_{(2)\mu}+KU_{(1)r}\,,
\end{equation}
so that
\begin{equation}
S_{\text{GH}}=\alpha\int_{\partial\mathcal{M}}d^2x\,e\left(\left(\hat e^\mu U_{(1)\mu}\right)^2-2\hat e^\mu U_{(1)\mu}\hat e^\nu\mathcal{L}_v\hat\tau_\nu\right)+\mathcal{O}(r^{-1})\,,
\end{equation}
where we wrote
\begin{equation}
\frac{1}{2}\epsilon_{\mu\nu}dx^\mu\wedge dx^\nu=d^2x\,e\,,
\end{equation}
and where we used that $v^\mu U_{(2)\mu}+KU_{(1)r}$ is a total derivative as follows from \eqref{eq:U1}, \eqref{eq:U2}, \eqref{eq:conditionU1} and \eqref{eq:K}.} $S_{\text{GH}}$ is order $\mathcal{O}(1)$. It can also be checked that if we first vary the GH boundary term and then go on-shell we find terms that are of order $r^0$, i.e. $\delta S_{\text{GH}}=O(1)$ on-shell.

Let us compute $U_Q J^Q$ at leading order. Using the results of appendix \ref{app:asymptotics} we find 
\begin{equation}
U_QJ^Q=r\left(e^{-1}\partial_\mu\left(e\delta v^\mu\right)-2\delta K-3U_{(1)\mu}\delta v^\mu\right)+\mathcal{O}(1)\,,
\end{equation}
where
\begin{equation}
\delta K=e^{-1}\partial_\mu\left[e\left(-\delta v^\mu-v^\mu e^{-1}\delta e\right)\right]-K e^{-1}\delta e\,.
\end{equation}
Using that $K=-v^\mu U_{(1)\mu}$ and that $U_{(1)\mu}=\partial_\mu U_{(1)}$ this can also be written as
\begin{eqnarray}
U_QJ^Q & = & r\left(e^{-1}\partial_\mu\left[ev^\mu\left(3\delta U_{(1)}-e^{-1}\delta e\right)\right]+K\left(3\delta U_{(1)}-e^{-1}\delta e\right)\right)+\mathcal{O}(1)\,,\nonumber\\
& = & rv^\mu\partial_\mu\left(3\delta U_{(1)}-e^{-1}\delta e\right)+\mathcal{O}(1)\,.
\end{eqnarray}
Since the variation of the GH boundary term is $\mathcal{O}(1)$ there is no way of canceling the variation with respect to $U_{(1)}$. It has to come from a boundary term involving the normal vector and the only candidate is the GH boundary term. We thus conclude that we must take\footnote{We could also take $U_{(1)}=\tfrac{1}{3}\log e+g$ where $g$ is some function such that $v^\mu\partial_\mu \delta g=0$, but we will not consider this case further.}
\begin{equation}
K=0\,,
\end{equation}
in order that $\delta S$ is of $\mathcal{O}(1)$ on-shell up to total derivative terms.

We continue by computing $U_QJ^Q$ up to $\mathcal{O}(1)$ for $K=0$. Let us write
\begin{equation}
U_QJ^Q=rX_{(1)}+X_{(2)}+\mathcal{O}(r^{-1})\,,
\end{equation}
where 
\begin{equation}
X_{(1)}=-e^{-1}\partial_\mu\left(ev^\mu e^{-1}\delta e\right)\,.
\end{equation}
The object of interest is the integrand of \eqref{eq:variationaction},
\begin{equation}
-\frac{1}{2}\epsilon_{MNP}dx^M\wedge dx^N V^P U_QJ^Q\vert_{\partial\mathcal{M}}=-ed^2x\left(Y_{(2)}+\mathcal{O}(r^{-1})\right)\,,
\end{equation}
where $Y_{(2)}$ is given by
\begin{equation}
Y_{(2)}=X_{(2)}+v^\mu\partial_\mu\left[\frac{1}{2}\hat h^{\rho\sigma}h_{(1)\rho\sigma}-v^\rho\partial_\rho\bar\Phi-U_{(1)r}\right] e^{-1}\delta e\,,
\end{equation}
where we ignored total derivative terms, used \eqref{eq:expintmeasure} as well as $\frac{1}{2}\epsilon_{\mu\nu}dx^\mu\wedge dx^\nu=ed^2x$. The task is to compute $X_{(2)}$ and subsequently $Y_{(2)}$. A lengthy calculation gives
\begin{equation}
Y_{(2)}=T_\mu\delta v^\mu-\frac{1}{2}T_{\mu\nu}\delta \hat h^{\mu\nu}\,,
\end{equation}
where 
\begin{eqnarray}
T_\mu & = & -e_\mu\left(\hat e^\nu\mathcal{L}_v\hat\tau_\nu\right)\hat h^{\rho\sigma}h_{(1)\rho\sigma}+e_\mu\left(\hat e^\nu\mathcal{L}_v\hat\tau_\nu\right)v^\rho\partial_\rho\bar\Phi+e_\mu\hat e^\nu\partial_\nu\left(\hat e^\rho\mathcal{L}_v\hat\tau_\rho\right)\nonumber\\
&&+e_\mu v^\nu\partial_\nu\left(\hat e^\rho h_{(1)r\rho}\right)+2e_\mu\hat e^\nu v^\rho h_{(2)\nu\rho}-\frac{1}{2}\partial_\mu\left(\hat h^{\rho\sigma}h_{(1)\rho\sigma}\right)-2\hat\tau_\mu\left(\hat e^\nu\mathcal{L}_v\hat\tau_\nu\right)^2\nonumber\\
&&-2\hat\tau_\mu\hat e^\nu\partial_\nu\left(\hat e^\rho\mathcal{L}_v\hat\tau_\rho\right)+\frac{1}{2}\hat\tau_\mu v^\nu\partial_\nu\left(\hat h^{\rho\sigma}h_{(1)\rho\sigma}\right)-3U_{(2)\mu}-3U_{(1)\mu}v^\nu\partial_\nu\bar\Phi\nonumber\\
&&+\frac{3}{2}U_{(1)\mu}\hat h^{\rho\sigma}h_{(1)\rho\sigma}-2\hat\tau_\mu v^\nu U_{(2)\nu}\,,\label{eq:Tmu}\\
T_{\mu\nu} & = & \left(-\frac{1}{2}v^\kappa\partial_\kappa\left(\hat h^{\rho\sigma}h_{(1)\rho\sigma}\right)-2v^\rho U_{(2)\rho}\right)h_{\mu\nu}\,.\label{eq:Tmunu}
\end{eqnarray}

For on-shell variations that respect $K=0$ we thus have that the variation of the bulk action without GH boundary term leads to
\begin{equation}\label{eq:osvariation}
\delta S\vert_{\text{os}}=\int_{\partial\mathcal{M}}d^2x e\left(-T_\mu\delta v^\mu+\frac{1}{2}T_{\mu\nu}\delta\hat h^{\mu\nu}\right)\,.
\end{equation}
Hence setting the variations $\delta v^\mu$ and $\delta\hat h^{\mu\nu}$ to zero at $\partial\mathcal{M}=\mathcal{I}^+$ we obtain a well-posed variational problem. 


\subsection{Ward identities}

In the variation \eqref{eq:osvariation} we need to ensure that we respect the condition $K=0$. From equation \eqref{eq:K} we learn that this is equivalent to 
\begin{equation}
\partial_\mu e_\nu-\partial_\nu e_\mu=0\,,
\end{equation}
so that $e_\mu=\partial_\mu f$ for some scalar function $f$. Using that
\begin{eqnarray}
\delta v^\mu & = & v^\mu v^\rho\delta\hat\tau_\rho-\hat e^\mu v^\rho\delta e_\rho\,,\\
\delta\hat h^{\mu\nu} & = & \left(v^\mu\hat e^\nu+v^\nu\hat e^\mu\right)\hat e^\rho\delta\hat\tau_\rho-\hat h^{\mu\nu}\hat h^{\rho\sigma}\delta h_{\rho\sigma}\,,
\end{eqnarray}
we can write upon partial integration
\begin{equation}\label{eq:osvariation2}
\delta S\vert_{\text{os}}=\int_{\partial\mathcal{M}}d^2x e\left[\left(v^\mu T_\mu v^\rho-v^\mu\hat e^\nu T_{\mu\nu}\hat e^\rho\right)\delta\hat\tau_\rho-e^{-1}\partial_\rho\left[e\left(\hat h^{\mu\nu}T_{\mu\nu}\hat e^\rho-\hat e^\mu T_\mu v^\rho\right)\right]\delta f\right]\,.
\end{equation}

Demanding invariance of the on-shell action under the local transformations \eqref{eq:localtrafosources1} and \eqref{eq:localtrafosources2} acting on the sources we obtain the following Ward identity due to the local transformation with parameter $e_\mu\chi_{(1)}^\mu$,
\begin{equation}\label{eq:chiWI}
v^\mu \hat e^\nu T_{\mu\nu}=0\,.
\end{equation}
We note that due to the fact that $\hat h^{\mu\nu}$ has one zero eigenvalue with eigenvector $\hat\tau_\mu$ the quantity $v^\mu v^\nu T_{\mu\nu}$ is not determined by \eqref{eq:osvariation}. In \eqref{eq:Tmunu} we chose $v^\mu v^\nu T_{\mu\nu}=0$ by hand. This is harmless as this quantity does not appear in any of the Ward identities. Next demanding that the on-shell action is invariant under the boundary diffeomorphisms $\delta\hat\tau_\rho=\mathcal{L}_\xi\hat\tau_\rho$ and $\delta f=\mathcal{L}_\xi f$ we obtain the diffeomorphism Ward identities
\begin{eqnarray}
0 & = & \mathcal{L}_v\left(v^\mu T_\mu\right)\,,\label{eq:WIA}\\
0 & = & \hat e^\rho\partial_\rho\left(\hat h^{\mu\nu}T_{\mu\nu}\right)+\hat e^\rho\mathcal{L}_v\hat\tau_\rho\left(\hat h^{\mu\nu}T_{\mu\nu}+v^\mu T_\mu\right)-\mathcal{L}_v\left(\hat e^\mu T_\mu\right)\,.\label{eq:WIB}
\end{eqnarray}
Equations \eqref{eq:WIA} and \eqref{eq:WIB} can also be written covariantly as
\begin{equation}\label{eq:covWI}
\overset{c}{\nabla}_\mu\mathcal{T}^\mu{}_\nu-2\overset{c}{\Gamma}{}^\mu_{[\mu\rho]}\mathcal{T}^\rho{}_\nu+2\overset{c}{\Gamma}{}^\rho_{[\mu\nu]}\mathcal{T}^\mu{}_\rho=0\,,
\end{equation}
where we defined 
\begin{equation}
\mathcal{T}^\mu{}_\nu=\hat h^{\mu\sigma}T_{\sigma\nu}-v^\mu T_\nu\,,
\end{equation}
and where we used the results of appendix \ref{app:Carrollgeometry}. 

Finally, the local transformations of the sources \eqref{eq:localtrafosources1} and \eqref{eq:localtrafosources2} also involve local dilatations with parameter $\Lambda_D$. However the condition $K=0$ forces the local dilatations to be such that $v^\mu\partial_\mu\Lambda_D=0$ in order that $\delta_D K=0$. A similar condition on Weyl transformations was observed in the context of asymptotically $z=2$ Schr\"odinger space-times that follow from an asymptotically AdS space-time by a TsT transformation \cite{Hartong:2010ec}. The invariance of the on-shell action under these restricted Weyl transformations only tells us that $-T_\mu v^\mu+T_{\mu\nu}\hat h^{\mu\nu}=v^\mu\partial_\mu G$ where $G$ is any function. This is compatible with the fact that solving the bulk equations of motion in section \ref{subsec:N2LO} did not give rise to a Ward identity for local Weyl invariance.

We will now match these equations with \eqref{eq:WI1} and \eqref{eq:WI2}. As we will see this requires us to choose an appropriate normal vector $U_M$. Equation \eqref{eq:WIA} is already of the form \eqref{eq:WI1}. In order to write \eqref{eq:WIB} in the form \eqref{eq:WI2} we use the following identity
\begin{equation}\label{eq:identity}
\hat e^\mu\partial_\mu\left(v^\nu\partial_\nu\left(\hat h^{\rho\sigma}h_{(1)\rho\sigma}\right)\right)+\hat e^\mu\mathcal{L}_v\hat\tau_\mu v^\nu\partial_\nu\left(\hat h^{\rho\sigma}h_{(1)\rho\sigma}\right)-\mathcal{L}_v\left(\hat e^\mu\partial_\mu\left(\hat h^{\rho\sigma}h_{(1)\rho\sigma}\right)\right)=0\,.
\end{equation}
Multiplying \eqref{eq:WIB} by $-1/2$ and adding $1/4$ times \eqref{eq:identity} we find using \eqref{eq:Tmu} and \eqref{eq:Tmunu}
\begin{eqnarray}
0 & = & \hat e^\rho\partial_\rho\left(-\frac{1}{2}\left(\hat e^\mu\mathcal{L}_v\hat\tau_\mu\right)^2+\frac{1}{2}v^\mu\partial_\mu\left(\hat h^{\lambda\kappa}h_{(1)\lambda\kappa}\right)+v^\mu U_{(2)\mu}\right)\nonumber\\
&&+2\hat e^\rho\mathcal{L}_v\hat\tau_\rho\left(-\frac{1}{2}\left(\hat e^\mu\mathcal{L}_v\hat\tau_\mu\right)^2+\frac{1}{2}v^\mu\partial_\mu\left(\hat h^{\lambda\kappa}h_{(1)\lambda\kappa}\right)+\frac{3}{4}v^\mu U_{(2)\mu}\right)\nonumber\\
&&+\mathcal{L}_v\left(\frac{1}{2}\hat e^\mu T_\mu-\frac{1}{4}\hat e^\mu\partial_\mu\left(\hat h^{\rho\sigma}h_{(1)\rho\sigma}\right)\right)\,.
\end{eqnarray}
Equation \eqref{eq:conditionU2} tells us that for $K=0$,
\begin{equation}
\frac{1}{2}v^\mu v^\nu h_{(2)\mu\nu}=v^\mu U_{(2)\mu}+\frac{1}{2}\left(\hat e^\mu U_{(1)\mu}-\hat e^\mu\mathcal{L}_v\hat\tau_\mu\right)^2 \,.
\end{equation}
We thus see that the diffeomorphism Ward identities \eqref{eq:WIA} and \eqref{eq:WIB} match \eqref{eq:WI1} and \eqref{eq:WI2} if and only if we have
\begin{eqnarray}
0 & = & \mathcal{L}_v\left(\frac{3}{2}v^\mu U_{(2)\mu}+\frac{1}{2}\left(\hat e^\mu U_{(1)\mu}-\hat e^\mu\mathcal{L}_v\hat\tau_\mu\right)^2-\frac{1}{2}\left(\hat e^\mu\mathcal{L}_v\hat\tau_\mu\right)^2\right)\,,\label{eq:matching1}\\
0 & = & \left(\hat e^\rho\partial_\rho+2\hat e^\rho\mathcal{L}_v\hat\tau_\rho\right)\left(v^\mu U_{(2)\mu}+\frac{1}{2}\left(\hat e^\mu U_{(1)\mu}-\hat e^\mu\mathcal{L}_v\hat\tau_\mu\right)^2-\frac{1}{2}\left(\hat e^\mu\mathcal{L}_v\hat\tau_\mu\right)^2\right)\nonumber\\
&&-\frac{3}{2}\hat e^\rho U_{(1)\rho}\mathcal{L}_v\left(v^\mu\partial_\mu\bar\Phi-\frac{1}{2}\hat h^{\mu\nu}h_{(1)\mu\nu}+U_{(1)r}\right)-\frac{3}{2}\hat e^\mu U_{(1)\mu}v^\nu U_{(2)\nu}\nonumber\\
&&-\frac{1}{2}\hat e^\mu\partial_\mu\left(v^\nu U_{(2)\nu}\right)\,.\label{eq:matching2}
\end{eqnarray}
Using the integrability condition for $U_{(1)\mu}$, equation \eqref{eq:intbU1}, i.e. $\mathcal{L}_v\left(\hat e^\mu U_{(1)\mu}\right)=0$ we can write \eqref{eq:matching1} as
\begin{equation}\label{eq:matching1a}
\frac{3}{2}\mathcal{L}_v\left(v^\mu U_{(2)\mu}\right)-\hat e^\mu U_{(1)\mu}\mathcal{L}_v\left(\hat e^\nu\mathcal{L}_v\hat\tau_\nu\right)=0\,.
\end{equation}
The equations \eqref{eq:matching2} and \eqref{eq:matching1a} tell us what the right choice of $U_M$ is. We will refer to these equations as the matching equations as they match the Ward identities \eqref{eq:WIA} and \eqref{eq:WIB} with the on-shell relations \eqref{eq:WI1} and \eqref{eq:WI2}. We will comment on their significance in the next section where we study the most general solution dual to a torsion free boundary geometry\footnote{We recall that all 1+1 dimensional boundary geometries are flat Carrollian space-times with torsion, so a special circumstance is to be torsion free.}. 

We started our analysis with a completely general hypersurface orthogonal null vector $U_M$ and now we see that in order for holography to work we need certain conditions on $U_M$ to be obeyed. This simply means that the family of hypersurfaces $F$ in \eqref{eq:hypersurfaces} contains more than just $\mathcal{I}^+$ for large $r$ and we need extra conditions to single out the right normal vector $U_M$. In AdS we would never have such a scenario because there we know that the Fefferman--Graham holographic coordinate $r$ not only makes local Weyl transformations manifest it also describes the direction normal to the boundary. In asymptotically flat space-times in BMS gauge the holographic coordinate $r$ makes the $z=1$ local dilatations manifest but not the boundary.

\section{Torsion free boundaries and BMS symmetries}\label{sec:cstsources}

In this section we take a closer look at the boundary geometries for which the sources are $\bar\Phi=0$, $\hat\tau_\mu=\delta_\mu^u$ and $h_{\mu\nu}=\delta_\mu^\varphi\delta_\nu^\varphi$. These are torsion free Carrollian geometries and we will see that the diffeomorphism Ward identity gives rise BMS symmetries and associated conserved currents. Further we will study the matching equations for this class of boundary geometries and show that one of them agrees with the condition $\partial_u^2h_{(1)\varphi\varphi}=0$ imposed in \cite{Bagchi:2012yk}. The other matching equation merely restricts the form of the normal vector $U_M$.

\subsection{Most general solution}\label{subsec:flatPBHtrafos}

We can always gauge fix bulk diffeomorphisms such that
\begin{equation}\label{eq:flatbdry}
\bar\Phi=0\,,\qquad\hat\tau_\mu=\delta_\mu^u\,,\qquad h_{\mu\nu}=\delta_\mu^\varphi\delta_\nu^\varphi\,.
\end{equation}
Demanding that 
\begin{equation}
\delta\bar\Phi=0\,,\qquad\delta\hat\tau_\mu=0\,,\qquad \delta h_{\mu\nu}=0\,,
\end{equation}
we find the following coordinate transformations that leave the sources invariant,
\begin{eqnarray}
\chi^u & = & uf'(\varphi)+g(\varphi)\,,\label{eq:residual1}\\
\chi^\varphi & = & f(\varphi)\,,\label{eq:residual2}\\
\chi^u_{(1)} & = & 0\,,\\
\chi^\varphi_{(1)} & = & -uf''(\varphi)-g'(\varphi)\,,\\
\Lambda_D & = & -f'(\varphi)\,,\label{eq:residual5}
\end{eqnarray}
where $f$ and $g$ are arbitrary functions of $\varphi$. Considering subleading orders we can use $\chi_{(2)}^\mu$ to set $h_{(1)rr}=h_{(1)r\varphi}=0$. We can continue this way and show that we can set $g_{rr}=0$ and $g_{r\mu}=-\delta_\mu^u$ exactly to all orders while having a flat boundary. The $r$ expansion then terminates and we find the following exact solution
\begin{equation}\label{eq:flatbdrysolution}
ds^2=-2dudr+h_{(2)uu}du^2+2h_{(2)u\varphi}dud\varphi+\left(r+\frac{1}{2}h_{(1)\varphi\varphi}\right)^2d\varphi^2\,,
\end{equation}
where 
\begin{eqnarray}
\partial_u\left( h_{(2)uu}+\partial_u h_{(1)\varphi\varphi}\right) & = & 0\,,\label{eq:flatWI1}\\
\partial_\varphi h_{(2)uu}-2\partial_u h_{(2)u\varphi} & = & 0\,.\label{eq:flatWI2}
\end{eqnarray}
Any other solution is related to this one by a coordinate transformation.

It can be shown that under the residual coordinate transformations
\begin{eqnarray}
\delta h_{(1)\varphi\varphi} & = & \chi^\rho\partial_\rho h_{(1)\varphi\varphi}+f'h_{(1)\varphi\varphi}-2\partial_\varphi^2\chi^u+2\xi_{(1)}^r\,,\label{eq:deltah1}\\
\delta h_{(2)uu} & = & \chi^\rho\partial_\rho h_{(2)uu}+2f'h_{(2)uu}-2\partial_u\xi_{(1)}^r\,,\label{eq:deltah2uu}\\
\delta h_{(2)u\varphi} & = & \chi^\rho\partial_\rho h_{(2)u\varphi}+2f'h_{(2)u\varphi}+\partial_\varphi\chi^uh_{(2)uu}-\frac{1}{2}f''h_{(1)\varphi\varphi}\nonumber\\
&&+\frac{1}{2}\partial_\varphi\chi^u\partial_u h_{(1)\varphi\varphi}-\partial_\varphi\xi_{(1)}^r\,,\label{eq:deltah2varphivarphi}
\end{eqnarray}
with $\delta r=-\xi^r$ and $\delta x^\mu=-\xi^\mu$ with $\xi^r$ and $\xi^\mu$ given by \eqref{eq:xir} and \eqref{eq:ximu} the metric \eqref{eq:flatbdrysolution}, \eqref{eq:flatWI1} and \eqref{eq:flatWI2} remains form invariant. We can use the coordinate transformation whose parameter is $\xi^r_{(1)}$ to set $h_{(1)\varphi\varphi}=0$ as in \cite{Barnich:2010eb,Barnich:2013yka}, but we will not do so. Note that we still have a fully unconstrained parameter $\xi_{(1)}^r$ at our disposal. It can be checked that \eqref{eq:deltah1}--\eqref{eq:deltah2varphivarphi} leave the equations \eqref{eq:flatWI1} and \eqref{eq:flatWI2} invariant.

\subsection{Matching equations and determining the normal vector}\label{subsec:matchingeqs}

What can we say about the normal vector at $\mathcal{I}^+$ for a metric in the gauge \eqref{eq:flatbdrysolution}? Using that $U_M$ is null we know via equations \eqref{eq:conditionU1} and \eqref{eq:conditionU2} that
\begin{eqnarray}
U_{(1)u} & = & 0\,,\label{eq:U1u}\\
-U_{(2)u}+\frac{1}{2}\left(U_{(1)\varphi}\right)^2 & = & \frac{1}{2}h_{(2)uu}\,,\label{eq:U2u}
\end{eqnarray}
where we used \eqref{eq:flatbdry}. Further since $U_M$ is hypersurface orthogonal we have through equations \eqref{eq:intbU1} and \eqref{eq:intbU2}
\begin{eqnarray}
\partial_u U_{(1)\varphi} & = & 0\,,\label{eq:1stintcond}\\
\partial_u U_{(2)\varphi} & = & \partial_\varphi U_{(2)u}+U_{(2)u}U_{(1)\varphi}\,.\label{eq:2ndintcond}
\end{eqnarray}
It is clear that these conditions alone do not fix what $U_M$ should be.

Next turning to the matching equations \eqref{eq:matching1a} and \eqref{eq:matching2} we find
\begin{eqnarray}
0 & = & \partial_u U_{(2)u}\,,\label{eq:matchingflat1}\\
0 & = & \partial_\varphi U_{(2)u}-2U_{(1)\varphi}\left(\partial_\varphi U_{(1)\varphi}-\frac{3}{4}\partial_u h_{(1)\varphi\varphi}+\frac{3}{2}U_{(2)u}\right)\,.\label{eq:matchingflat2}
\end{eqnarray}
It follows from \eqref{eq:U2u}, \eqref{eq:1stintcond} and \eqref{eq:matchingflat1} that $\partial_u h_{(2)uu}=0$. Hence the Ward identity \eqref{eq:flatWI1} becomes
\begin{equation}\label{eq:strongerWI}
\partial_u^2 h_{(1)\varphi\varphi}=0=\partial_u h_{(2)uu}\,,
\end{equation}
which is the condition imposed in \cite{Bagchi:2012yk}. The Ward identities obtained from demanding diffeomorphism invariance of the on-shell action at future null infinity are stronger than those written in \eqref{eq:flatWI1} and \eqref{eq:flatWI2}. For example we can obtain the more general version \eqref{eq:flatWI1} by starting with \eqref{eq:strongerWI} and demanding invariance under arbitrary $\xi_{(1)}^r$ transformations. It would be interesting to understand this and the role of the $\xi_{(1)}^r$ transformation better.

The necessary conditions $U_M$ has to obey are given in equations \eqref{eq:U1u}--\eqref{eq:matchingflat2}. We can remove $U_{(2)u}$ from these equations by using \eqref{eq:U2u} to express it in terms of $h_{(2)uu}$ and $U_{(1)\varphi}$. The second equation \eqref{eq:matchingflat2} then becomes a first order ordinary nonlinear differential equation for $U_{(1)\varphi}$ that reads
\begin{equation}\label{eq:diffeqU1}
2U_{(1)\varphi}U_{(1)\varphi}'-3U_{(1)\varphi}\partial_u h_{(1)\varphi\varphi}+3U_{(1)\varphi}\left(-h_{(2)uu}+U_{(1)\varphi}^2\right)+\partial_\varphi h_{(2)uu}=0\,,
\end{equation}
where a prime denotes differentiation with respect to $\varphi$. Given a solution for $U_{(1)\varphi}$ equation \eqref{eq:U2u} then tells us what $U_{(2)u}$ is. The component $U_{(2)\varphi}$ will not be fully determined except its $u$ derivative via equation \eqref{eq:2ndintcond}. The u-independent part can be removed by using the freedom to perform $\xi_{(1)}^r$ coordinate transformations.

We thus see that given the metric \eqref{eq:flatbdrysolution} the normal is almost fully determined. The only indeterminacy lies in the number of solutions to \eqref{eq:diffeqU1}. This is due to the fact that our methodology leading up to the matching equations only provides necessary conditions. To illustrate this note that for a constant $U_{(2)u}$, equation \eqref{eq:matchingflat2} leads to two possibilities, namely $U_{(1)\varphi}=0$ or $\partial_\varphi U_{(1)\varphi}-\frac{3}{4}\partial_u h_{(1)\varphi\varphi}+\frac{3}{2}U_{(2)u}=0$. It would be interesting to have an a priori understanding of which $U_{(1)\varphi}$ solution to \eqref{eq:diffeqU1} one should take. 

Consider the following simple example
\begin{equation}
ds^2=Cdu^2-2dudr+r^2d\varphi^2\,,
\end{equation}
where $h_{(2)uu}=C$ is a constant. For $C<0$ the metric describes a cone which for $C=-1$ is just Minkowski space-time. For $C>0$ we are dealing with compactified Milne space-time while for $C=0$ we have a null cone. Equations \eqref{eq:U1u}--\eqref{eq:2ndintcond} apply and the matching equation \eqref{eq:matchingflat2} becomes
\begin{equation}
0 = U_{(1)\varphi}\left[\partial_\varphi U_{(1)\varphi}+\frac{3}{2}\left(U_{(1)\varphi}\right)^2  -\frac{3}{2}C\right]\,.
\end{equation}
We will take the solution $U_{(1)\varphi}=0$. It then follows that $U_{(2)u}=-C/2$ where we used \eqref{eq:U2u}. The component $U_{(2)\varphi}$ is then a function of $\varphi$ that can be set equal to zero by using the $\xi_{(1)}^r$ transformation. We thus find the following hypersurface $F=r-C/2 u$ as future null infinity.

Using \eqref{eq:trafoU1} and \eqref{eq:trafoU2} we see that under the residual coordinate transformations \eqref{eq:residual1}--\eqref{eq:residual5} the normal vector transforms as 
\begin{eqnarray}
\delta U_{(1)\varphi} & = & f\partial_\varphi U_{(1)\varphi}+f'U_{(1)\varphi}-f''\,,\\
\delta U_{(2)u} & = & f\partial_\varphi U_{(2)u}+2f'U_{(2)u}-f'' U_{(1)\varphi}+\partial_u\xi_{(1)}^r\,,\\
\delta U_{(2)\varphi} & = & \chi^\rho\partial_\rho U_{(2)\varphi}+2f'U_{(2)\varphi}+\partial_\varphi\chi^u U_{(2)u}+\left(\xi_{(1)}^r+\alpha_{(1)}\right)U_{(1)\varphi}\nonumber\\
&&-\partial_\varphi\chi^u\partial_\varphi U_{(1)\varphi}-U_{(1)\varphi}\partial_\varphi^2\chi^u-f'' U_{(1)r}+\partial_\varphi\xi_{(1)}^r\,,
\end{eqnarray}
where $\chi^\mu$ is given by \eqref{eq:residual1} and \eqref{eq:residual2}. The metric \eqref{eq:flatbdrysolution} is form invariant under \eqref{eq:deltah1}--\eqref{eq:deltah2varphivarphi} and the corresponding transformation of the coordinates. This is of course also true for the relations \eqref{eq:flatWI1} and \eqref{eq:flatWI2}. Further the conditions \eqref{eq:U1u}--\eqref{eq:2ndintcond} are form invariant as well, but the matching equations are in general not unless we impose the following conditions on $\xi_{(1)}^r$
\begin{eqnarray}
\partial_u^2\xi_{(1)}^r & = & 0\,,\\
\partial_u\partial_\varphi\xi_{(1)}^r & = & 2U_{(1)\varphi}f'''-f''\left(\partial_\varphi U_{(1)\varphi}-\frac{3}{2}\partial_u h_{(1)\varphi\varphi}+5U_{(2)u}+U_{(1)\varphi}^2\right)\,,
\end{eqnarray}
where we used $U_{(1)r}=0$ and $\alpha_{(1)}=-U_{(1)\varphi}\partial_\varphi\chi^u$ so that $\delta U_{(1)r}=0$. The parameter $\xi_{(1)}^r$ is thus determined up to an arbitrary function of $\varphi$. The restrictions we find on $\xi_{(1)}^r$ are related to the comments made in the last two paragraphs of the previous subsection where we showed that the Ward identities obtained from the diffeomorphism invariance of the on-shell action are not invariant under generic $\xi_{(1)}^r$ transformations.

\subsection{BMS symmetries}\label{subsec:BMS}

If we contract \eqref{eq:covWI} with a vector $K^\mu$ we can write the result as
\begin{equation}
e^{-1}\partial_\rho\left[eK^\mu\left(\hat h^{\rho\nu}T_{\mu\nu}-v^\rho T_\mu\right)\right]-T_\mu\mathcal{L}_K v^\mu+\frac{1}{2}T_{\mu\nu}\mathcal{L}_K\hat h^{\mu\nu}=0\,.
\end{equation}
Hence for any solution to the Killing equations
\begin{equation}
\mathcal{L}_K v^\mu=0\,,\qquad\mathcal{L}_K \hat h^{\mu\nu}=0\,.
\end{equation}
we have a conserved current. More is true, since we have 
\begin{eqnarray}
v^\mu T_\mu & = & e^{-1}\partial_\rho\left(-ev^\rho\hat h^{\mu\nu}h_{(1)\mu\nu}+2e\hat e^\rho\hat e^\mu\mathcal{L}_v\hat\tau_\mu\right)-v^\mu U_{(2)\mu}\,,\\
\hat h^{\mu\nu} T_{\mu\nu} & = & -\frac{1}{2}e^{-1}\partial_\rho\left(ev^\rho\hat h^{\mu\nu}h_{(1)\mu\nu}\right)-2v^\mu U_{(2)\mu}\,,
\end{eqnarray}
it follows that for any solution to 
\begin{equation}\label{eq:moregenK}
\mathcal{L}_K v^\mu=\Omega v^\mu\,,\qquad\mathcal{L}_K \hat h^{\mu\nu}=2\Omega \hat h^{\mu\nu}-\zeta^\mu v^\nu-\zeta^\nu v^\mu\,,
\end{equation}
where $\Omega$ is constant we have the conservation equation
\begin{equation}
\partial_\rho\left(e\mathcal{J}^\rho\right)=0\,,
\end{equation}
where the current $\mathcal{J}^\rho$ is given by
\begin{equation}\label{eq:conservedcurrents}
\mathcal{J}^\rho=K^\mu\mathcal{T}^\rho{}_\mu-2\Omega\hat e^\rho\hat e^\mu\mathcal{L}_v\hat\tau_\mu+\Omega v^\rho\left(U_{(1)r}-e^{-U_{(1)}}U_{(2)}+\frac{1}{2}\hat h^{\mu\nu}h_{(1)\mu\nu}\right)\,.
\end{equation}
We used that $v^\mu T_{\mu\nu}=0$ so that the $\zeta^\mu$ drops out. The form of the right hand side of \eqref{eq:moregenK} is such that the vector $K^\mu$ corresponds to a boundary diffeomorphism that is generated from a bulk diffeomorphism leaving invariant the sources, i.e. setting to zero the variations in \eqref{eq:sourcetrafo1} and \eqref{eq:sourcetrafo2}. When $\hat e^\mu\mathcal{L}_v\hat\tau_\mu=0$ there is no torsion as then \eqref{eq:torsion} vanishes (using that $K=0$). In this case we can have a non-constant $\Omega$ if it obeys $v^\mu\partial_\mu\Omega=0$. We call vectors $K^\mu$ obeying \eqref{eq:moregenK} generalized conformal Killing vectors, where generalized refers to the presence of the $\zeta^\mu$ vector.

Let us consider the torsion free boundary of \eqref{eq:flatbdry}, so that $v^\mu\partial_\mu\Omega=0$. The most general solution to \eqref{eq:moregenK} is given by the Killing vectors
\begin{eqnarray}
K^\varphi & = & f(\varphi)\,,\\
K^u & = & f'(\varphi)u+g(\varphi)\,,\\
\Omega & = & -f'(\varphi)\,,\\
\zeta^u & = & 0\,,\\
\zeta^\varphi & = & -f''(\varphi)u-g'(\varphi)\,,
\end{eqnarray}
which are of course of the same form as the residual coordinate transformations \eqref{eq:residual1}--\eqref{eq:residual5}. 

For the existence of this infinite dimensional symmetry algebra it is crucial that $\zeta^\mu$ is nonzero. The fact that $\zeta^\varphi$ is nonzero is related to the $e_\mu\chi_{(1)}^\mu$ diffeomorphisms and the corresponding Ward identity \eqref{eq:chiWI} which can be written as $e_\mu v^\nu\mathcal{T}^\mu{}_\nu=0$. For \eqref{eq:flatbdry} this becomes $\mathcal{T}^\varphi{}_u=0$. Since the sources are constant the affine connection \eqref{eq:affineconnection} vanishes so that the diffeomorphism Ward identity \eqref{eq:covWI} becomes\footnote{Similar expressions have been found by contraction of the AdS$_3$ boundary energy-momentum tensor in \cite{Fareghbal:2013ifa}.}
\begin{eqnarray}
\partial_u \mathcal{T}^u{}_u+\partial_\varphi \mathcal{T}^\varphi{}_u & = & 0\,,\\
\partial_u \mathcal{T}^u{}_\varphi+\partial_\varphi \mathcal{T}^\varphi{}_\varphi & = & 0\,.
\end{eqnarray}
Hence $\mathcal{T}^\varphi{}_u$ is like an energy flux which in a BMS$_3$ invariant theory must vanish\footnote{If we interchange space and time, i.e. replace $e_\mu$ by the Newton--Cartan clock 1-form $\tau_\mu^{\text{NC}}$ and $\tau_\mu$ by the Newton--Cartan spatial vielbein $e_\mu^{\text{NC}}$ we would be dealing with a geometry that can be thought of as arising from gauging the massless Galilean algebra (that is without the Bargmann extention). The corresponding Ward identity would tell us that the momentum current vanishes which makes perfect sense since the Galilean boost Ward identity \cite{Jensen:2014aia,Hartong:2014pma} relates that current to the mass current which is zero since we are dealing with massless Galilean theories.}. Further we have for the other components
\begin{eqnarray}
\mathcal{T}^u{}_u & = & -\partial_u h_{(1)\varphi\varphi}-U_{(2)u}\,,\label{eq:EMT1}\\
\mathcal{T}^u{}_\varphi & = & -2h_{(2)u\varphi}-\frac{1}{2}\partial_\varphi h_{(1)\varphi\varphi}-3U_{(2)\varphi}+\frac{3}{2}U_{(1)\varphi}h_{(1)\varphi\varphi}\,,\\
\mathcal{T}^\varphi{}_\varphi & = & \frac{1}{2}\partial_u h_{(1)\varphi\varphi}+2U_{(2)u}\,.\label{eq:EMT3}
\end{eqnarray}

The coordinate $\varphi$ is periodic with period $2\pi$. Let us write instead of the most general Killing vector $K$ the Killing vectors $L$ and $M$ defined as
\begin{eqnarray}
L & = & f(\varphi)\partial_\varphi+f' u\partial_u\,,\\
M & = & g(\varphi)\partial_u\,,
\end{eqnarray}
so that $K=L+M$. We Fourier decompose $f$ and $g$ as
\begin{eqnarray}
f(\varphi) & = & \sum_{n=-\infty}^\infty a_{n} e^{in\varphi}\,,\label{eq:Fourierf}\\
g(\varphi) & = & \sum_{n=-\infty}^\infty b_{n} e^{in\varphi}\,,\label{eq:Fourierg}
\end{eqnarray}
where $a^*_n=a_{-n}$ and $b^*_n=b_{-n}$ for reality and we define the complex coordinate $z$ as $z=e^{i\varphi}$. We then have $\partial_\varphi=iz\partial_z$. Defining $L_n$ and $M_n$ via
\begin{eqnarray}
L & = & -i\sum_{n=-\infty}^\infty a_{n} L_{n}\,,\\
M & = & \sum_{n=-\infty}^\infty b_{n} M_{n}\,,
\end{eqnarray}
we obtain\footnote{To write the generators in a manner that is compatible with say \cite{Bagchi:2009my} one can make the following redefinitions $L_{n+1}=z\tilde L_n$ and $M_{n+1}=\tilde M_n$. The tilded generators are then given by $\tilde L_n = -z^{n+1}\partial_z-(n+1)z^nu\partial_u$ and $\tilde M_n = z^{n+1}\partial_u$ that satisfy the same algebra \eqref{eq:LL}--\eqref{eq:LM}.}
\begin{eqnarray}
L_n & = & -z^{n+1}\partial_z-nz^nu\partial_u\,,\\
M_n & = & z^{n}\partial_u\,.
\end{eqnarray}
The generators $L_n$ and $M_n$ span the BMS$_3$ algebra
\begin{eqnarray}
\left[L_m\,,L_n\right] & = & (m-n)L_{m+n}\,,\label{eq:LL}\\
\left[M_m\,,M_n\right] & = & 0\,,\label{eq:MM}\\
\left[L_m\,,M_n\right] & = & (m-n)M_{m+n}\,.\label{eq:LM}
\end{eqnarray}

The BMS$_3$ currents \eqref{eq:conservedcurrents} for the solution \eqref{eq:flatbdrysolution} are given by
\begin{eqnarray}
\mathcal{J}^u & = & \left(f'u+g\right)\mathcal{T}^u{}_u+f\mathcal{T}^u{}_\varphi+f'\left(-e^{-U_{(1)}}U_{(2)}+\frac{1}{2}h_{(1)\varphi\varphi}\right)\,,\\
\mathcal{J}^\varphi & = & f\mathcal{T}^\varphi{}_\varphi\,,
\end{eqnarray}
where the components of the energy-momentum tensor are given by \eqref{eq:EMT1}--\eqref{eq:EMT3}. The conservation equation is just
\begin{equation}
\partial_u \mathcal{J}^u+\partial_\varphi \mathcal{J}^\varphi=0\,.
\end{equation}
It can be checked that the terms containing the normal vector drop out of the divergence by using \eqref{eq:U1u}--\eqref{eq:matchingflat2}, i.e.
\begin{equation}
\partial_u \mathcal{J}^u+\partial_\varphi \mathcal{J}^\varphi=f\left(\partial_\varphi h_{(2)uu}-2\partial_u h_{(2)\varphi\varphi}\right)-g\partial_u^2h_{(1)\varphi\varphi}\,,
\end{equation}
which vanishes on account of \eqref{eq:flatWI2} and \eqref{eq:strongerWI}. We are setting $U_{(1)r}=0$. The $U_M$ dependent terms can be written in terms of a current that is conserved merely on the basis of the properties of the normal vector. More precisely if we define the current $\mathcal{I}^\mu$ as
\begin{eqnarray}
\mathcal{I}^u & = & -\left(f' u+g\right)\left(3U_{(2)u}-U_{(1)\varphi}^2\right)-3fU_{(2)\varphi}+\frac{3}{2}f U_{(1)\varphi}h_{(1)\varphi\varphi}-f' e^{-U_{(1)}}U_{(2)}\,,\\
\mathcal{I}^\varphi & = & 2fU_{(2)u}\,,
\end{eqnarray}
then we have
\begin{equation}
\partial_u \mathcal{I}^u+\partial_\varphi \mathcal{I}^\varphi=0\,,
\end{equation}
as a consequence of \eqref{eq:U1u}--\eqref{eq:matchingflat2}. We can thus define a new conserved current $\tilde{\mathcal{J}}^\mu=\mathcal{J}^\mu-\mathcal{I}^\mu$ whose components read
\begin{eqnarray}
\tilde{\mathcal{J}}^u & = & -\left(f'u+g\right)\left(\partial_u h_{(1)\varphi\varphi}+h_{(2)uu}\right)-2f h_{(2)u\varphi}+f' h_{(1)\varphi\varphi}-\frac{1}{2}\partial_\varphi\left(fh_{(1)\varphi\varphi}\right)\,,\\
\tilde{\mathcal{J}}^\varphi & = & \frac{1}{2}\partial_u\left(fh_{(1)\varphi\varphi}\right)\,.
\end{eqnarray}
The last terms in $\tilde{\mathcal{J}}^u$ together with the only term in $\tilde{\mathcal{J}}^\varphi$ form an identically conserved current. If we remove these pieces the $u$-component of $\tilde{\mathcal{J}}^\mu$ agrees with the integrands for the charges of the BMS$_3$ algebra as given in for example \cite{Bagchi:2012yk} where in order to compare we need to expand $f$ and $g$ into its Fourier modes and take the parameter $\mu$ in \cite{Bagchi:2012yk} to infinity in order to have the same bulk action. As shown in \cite{Barnich:2006av} the charge algebra gives rise to a central element in the $[L_m,M_n]$ commutator.

The solutions \eqref{eq:flatbdrysolution} contain interesting space-times such as conical defects \cite{Deser:1983tn} and cosmological solutions (orbifolds of flat space-time) \cite{Bagchi:2012xr}. The charges of these solutions can be computed using the expressions given in \cite{Bagchi:2012yk}.

\section{Discussion}

We have shown that the boundary geometry at $\mathcal{I}^+$ is described by the Carrollian geometry of \cite{Hartong:2015xda}. A covariant description of the boundary geometry allows for a definition of the boundary energy-momentum tensor $\mathcal{T}^\mu{}_\nu$ which satisfies two Ward identities: a diffeomorphism Ward identity and another one related to a shift invariance acting on the boundary source $\hat\tau_\mu$ which states that the energy flux $v^\nu e_\mu\mathcal{T}^\mu{}_\nu$ vanishes. It is this extra Ward identity that is responsible for the appearance of the infinite dimensional BMS$_3$ symmetry algebra. We showed that there is a well-posed variational problem at $\mathcal{I}^+$ without the need of adding a Gibbons--Hawking boundary term. The diffeomorphism Ward identity deriving from the diffeomorphism invariance of the on-shell action is compatible with the Ward identity-type equations obtained by solving Einstein's equations in the bulk.

This work can be extended in a number of ways. First of all it would be interesting to study the theory close to spatial infinity by working out the boundary geometry and using this to define a boundary energy-momentum tensor. Further, it would be interesting to compare the present techniques with the approach in \cite{Bagchi:2015wna} where a Chern--Simons formulation is used to compute correlation functions of stress tensor correlators.

The real interesting challenge though is to extend these methods to 4 space-time dimensions to make contact with black holes physics and S-matrix results \cite{Strominger:2013jfa,Kapec:2014opa}. One of the motivations for this work is to find an approach that does not use specific 3D techniques such as Chern--Simons theories or large radius AdS$_3$ limits in order to be able to study flat space holography in 4 dimensions. Of course there are many new features, notably the presence of gravitons, when going up in dimensions, but it would be interesting to see how far one can get by following similar reasoning. 

It is clearly important that we understand field theories (especially in 2 and 3 space-time dimensions) with BMS symmetries. Since we do not have a specific proposal for a duality between some quantum gravity theory on flat space-time and a theory on its boundary we need to resort to general characteristic features. A covariant description of the boundary geometry will help in this endeavor. In this work for example it allowed us to define a boundary energy-momentum tensor $\mathcal{T}^\mu{}_\nu$ and we showed that the energy flux $v^\nu e_\mu\mathcal{T}^\mu{}_\nu$ has to vanish in order to obtain all the BMS$_3$ currents. 

Finally, since any null hypersurface is a Carrollian geometry understanding field theory on Carrollian geometry in general might also be insightful for black hole physics in relation to the physics of black hole horizons. For example in relation to recent ideas concerning BMS supertranslations on black hole horizons \cite{Hawking:2015qqa}.

\section*{Acknowledgments}

I would like to thank the following people for many useful discussions: Arjun Bagchi, Glenn Barnich, Geoffrey Comp\`ere, St\'ephane Detournay, Gaston Giribet, Joaquim Gomis, Daniel Grumiller, Kristan Jensen, Francesco Nitti, Niels Obers, Jan Rosseel and Stefan Vandoren. Further, I would like to thank Stony Brook University, where this project was finalized, for its hospitality. The work of JH is supported by the advanced ERC grant `Symmetries and Dualities in Gravity and M-theory' of Marc Henneaux.

\appendix

\section{Carrollian Geometry}\label{app:Carrollgeometry}

In this appendix we review Carrollian geometry as defined in \cite{Hartong:2015xda} (see also \cite{Duval:2014uoa,Duval:2014uva,Bekaert:2015xua} for related work). In \cite{Hartong:2015xda} it is shown that one can gauge the Carroll algebra, deform it so as to replace local time and space translations by diffeomorphisms (in the same manner as was done for the gauging of the Poincar\'e and Galilei algebras in \cite{Hartong:2015zia}). This leads to a rather large set of independent fields, namely $\tau_\mu, e_\mu^a$ and two connections $\Omega_\mu{}^a$ and $\Omega_\mu{}^{ab}$ for local Carroll boosts and local spatial rotations respectively. One can then show that the algebra of diffeomorphisms and local tangent space transformations consisting of local Carroll boosts and local spatial rotations can also be realized on a smaller set of fields, i.e.  $\tau_\mu, e_\mu^a$ and a contravariant vector $M^\mu$. It is in terms of these variables that null hypersurfaces are described by Carrollian geometry \cite{Duval:2014uoa,Duval:2014uva,Bekaert:2015xua,Hartong:2015xda}. This also applies to the boundary at future null infinity (as shown in section \ref{subsec:bdrygeometry}). The fields $\tau_\mu, e_\mu^a$ and $M^\mu$ allow us to write down an expression for a (torsionful) affine connection $\overset{c}{\Gamma}{}^{\rho}_{\mu\nu}$ that is invariant under the tangent space transformations and that is metric compatible in the sense of $\overset{c}{\nabla}_\mu v^\nu=0$ and $\overset{c}{\nabla}_\mu h_{\nu\rho}=0$. This expression will be given further below.

\subsection{Arbitrary dimensions}

The Carroll algebra is given by
\begin{eqnarray*}
\left[J_{ab},P_c\right]&=&\delta_{ac}P_b-\delta_{bc}P_a\,,\\
\left[J_{ab},C_c\right]&=&\delta_{ac}C_b-\delta_{bc}C_a\,,\\
\left[J_{ab},J_{cd}\right]&=&\delta_{ac}J_{bd}-\delta_{ad}J_{bc}-\delta_{bc}J_{ad}+\delta_{bd}J_{ac}\,,\\
\left[P_a,C_b\right]&=&\delta_{ab}H\,,
\end{eqnarray*}
where $a=1,\ldots,d$ with $H$ denoting the Hamiltonian, $P_a$ spatial momenta, $J_{ab}$ spatial rotations and $C_a$ Carrollian boosts. We introduce a Lie algebra valued connection $\mathcal{A}_\mu$ via
\begin{equation}
\mathcal{A}_\mu=H\tau_\mu+P_a e^a_\mu+C_a\Omega_\mu{}^a+\frac{1}{2}J_{ab}\Omega_\mu{}^{ab}\,,
\end{equation}
where $\mu$ takes $d+1$ values, so that we work with a $(d+1)$-dimensional space-time. This connection transforms in the adjoint so that we have
\begin{equation}
\delta\mathcal{A}_\mu=\partial_\mu\Lambda+[A_\mu,\Lambda]\,.
\end{equation}
One can always write $\Lambda$ as
\begin{equation}
\Lambda=\xi^\mu\mathcal{A}_\mu+\Sigma\,,
\end{equation}
where $\Sigma$ is given by 
\begin{equation}
\Sigma=C_a\lambda^a+\frac{1}{2}J_{ab}\lambda^{ab}\,.
\end{equation}
The idea will be to think of $\xi^\mu$ as the generator of diffeomorphisms and $\Sigma$ as the tangent space transformations. To realize this idea we need to introduce a new local transformation denoted by $\bar\delta$ that is defined as
\begin{equation}
\bar\delta\mathcal{A}_\mu=\delta\mathcal{A}_\mu-\xi^\nu\mathcal{F}_{\mu\nu}=\mathcal{L}_\xi\mathcal{A}_\mu+\partial_\mu\Sigma+[A_\mu,\Sigma]\,,
\end{equation}
where $\mathcal{F}_{\mu\nu}$ is the Yang--Mills field strength
\begin{eqnarray}
\mathcal{F}_{\mu\nu} & = & \partial_\mu\mathcal{A}_\nu-\partial_\nu\mathcal{A}_\mu+[\mathcal{A}_\mu,\mathcal{A}_\nu]\nonumber\\
& = & H R_{\mu\nu}(H)+P_a R_{\mu\nu}{}^a(P)+C_a R_{\mu\nu}{}^a(C)+\frac{1}{2}J_{ab}R_{\mu\nu}{}^{ab}(J)\,.
\end{eqnarray}
From now on we will always work with the $\bar\delta$ transformations and drop the bar on $\delta$.
In components the transformations act as
\begin{eqnarray}
\delta\tau_\mu & = & \mathcal{L}_\xi\tau_\mu+e^a_\mu\lambda_a\,,\label{eq:deltatrafo1}\\
\delta e_\mu^a & = & \mathcal{L}_\xi e_\mu^a+\lambda^a{}_b e^b_\mu\,,\\
\delta\Omega_\mu{}^a & = & \mathcal{L}_\xi\Omega_\mu{}^a+\partial_\mu\lambda^a+\lambda^a{}_b\Omega_\mu{}^b-\lambda_b\Omega_\mu{}^{ab}\,,\\
\delta\Omega_\mu{}^{ab} & = & \mathcal{L}_\xi\Omega_\mu{}^{ab}+\partial_\mu\lambda^{ab}+\lambda^a{}_c\Omega_\mu{}^{cb}-\lambda^b{}_c\Omega_\mu{}^{ca}\,.\label{eq:deltatrafo4}
\end{eqnarray}
The Lie derivatives along $\xi^\mu$ correspond to diffeomorphisms whereas the local transformations with parameters $\lambda^a$ and $\lambda^{ab}=-\lambda^{ba}$ correspond to local tangent space transformations. The Carroll algebra can be obtained as the $c\rightarrow 0$ contraction of the Poincar\'e algebra. Hence the light cones have collapsed to a line. This shows up in the fact that there are no boost transformations acting on the spacelike vielbeins $e_\mu^a$. It also features dynamically in that a free Carroll particle does not move \cite{Bergshoeff:2014jla}. The component expressions for the field strengths are given by
\begin{eqnarray}
R_{\mu\nu}(H) & = & \partial_\mu\tau_\nu-\partial_\nu\tau_\mu+e_\mu^a\Omega_{\nu a}-e^a_\nu\Omega_{\mu a}\,,\\
R_{\mu\nu}{}^a(P) & = & \partial_\mu e_\nu^a-\partial_\nu e_\mu^a-\Omega_\mu{}^{ab}e_{\nu b}+\Omega_\nu{}^{ab} e_{\mu b}\,,\\
R_{\mu\nu}{}^a(C) & = & \partial_\mu\Omega_\nu{}^a-\partial_\nu\Omega_\mu{}^a-\Omega_\mu{}^{ab}\Omega_{\nu b}+\Omega_\nu{}^{ab} e_{\mu b}\,,\label{eq:Ccurv}\\
R_{\mu\nu}{}^{ab}(J) & = & \partial_\mu\Omega_\nu{}^{ab}-\partial_\nu\Omega_\mu{}^{ab}-\Omega_\mu{}^{ca}\Omega_\nu{}^b{}_c+\Omega_\nu{}^{ca}\Omega_\mu{}^b{}_c\,.
\end{eqnarray}

We next introduce vielbein postulates so that we can describe the properties of the curvatures in $\mathcal{F}_{\mu\nu}$ in terms of the Riemann curvature and torsion of an affine connection $\Gamma^\rho_{\mu\nu}$. The affine connection is invariant under the tangent space $\Sigma$ transformations. We denote by $\mathcal{D}_\mu$ the covariant derivative  that transforms covariantly under the $\delta$ transformations. It is given by
\begin{eqnarray}
\mathcal{D}_\mu\tau_\nu & = & \partial_\mu\tau_\nu-\overset{c}{\Gamma}{}^\rho_{\mu\nu}\tau_\rho-\Omega_{\mu a}e^a_\nu\,,\label{eq:covdertau}\\
\mathcal{D}_\mu e^a_\nu & = & \partial_\mu e_\nu^a-\overset{c}{\Gamma}{}^\rho_{\mu\nu} e^a_\rho-\Omega_\mu{}^a{}_b e^b_\nu\,.
\end{eqnarray}
We can now write the vielbein postulates simply as 
\begin{eqnarray}
\mathcal{D}_\mu\tau_\nu & = & 0\,,\label{eq:VP1}\\
\mathcal{D}_\mu e_\mu^a & = & 0\,.\label{eq:VP2}
\end{eqnarray}
These equations can be solved for $\Omega_{\mu a}$ and $\Omega_\mu{}^a{}_b$ in terms of $\Gamma^\rho_{\mu\nu}$ by contracting the vielbein postulates with the inverse vielbeins $v^\mu$ and $e^\mu_a$. The inverse are defined through
\begin{equation}
v^\mu\tau_\mu=-1\,,\qquad v^\mu e^a_\mu=0\,,\qquad e^\mu_a\tau_\mu=0\,,\qquad e^\mu_ae^b_\mu=\delta^b_a\,,
\end{equation}
and they transform under the $\delta$ transformation as
\begin{eqnarray}
\delta v^\mu & = & \mathcal{L}_\xi v^\mu\,,\\
\delta e^\mu_a & = & \mathcal{L}_\xi e^\mu_a+v^\mu\lambda_a+\lambda_a{}^b e^\mu_b\,.
\end{eqnarray}
The inverse vielbein postulates read
\begin{eqnarray}
\mathcal{D}_\mu v^\nu & = & \partial_\mu v^\nu+\overset{c}{\Gamma}{}^\nu_{\mu\rho} v^\rho=0\,,\label{eq:VP3}\\
\mathcal{D}_\mu e^\nu_a & = & \partial_\mu e^\nu_a+\overset{c}{\Gamma}{}^\nu_{\mu\rho}e^\rho_a-v^\nu\Omega_{\mu a}-\Omega_{\mu a}{}^b e^\nu_b=0\,.\label{eq:VP4}
\end{eqnarray}
The vielbein postulates ensure metric compatibility in the sense that $\overset{c}{\nabla}_\mu v^\nu=0$ and $\overset{c}{\nabla}_\mu h_{\nu\rho}=0$ where $v^\mu$ and $h_{\mu\nu}=\delta_{ab}e^a_\mu e^b_\nu$ are the Carrollian metric-like quantities (i.e. invariant under the tangent space transformations). Here $\overset{c}{\nabla}_\mu$ denotes the covariant derivative containing $\overset{c}{\Gamma}{}^\nu_{\mu\rho}$. We put a superscript $c$ to indicate that these are covariant in the sense of Carrollian geometry and to distinguish them from the bulk covariant derivative and Levi--Civit\`a connection. 

By isolating the affine connection and taking the antisymmetric part of \eqref{eq:covdertau}--\eqref{eq:VP2} we see that the torsion $\overset{c}{\Gamma}{}^\rho_{[\mu\nu]}$ can be related to the curvatures $R_{\mu\nu}(H)$ and $R_{\mu\nu}{}^a(P)$ via
\begin{equation}
2\overset{c}{\Gamma}{}^\rho_{[\mu\nu]}=-v^\rho R_{\mu\nu}(H)+e^\rho_a R_{\mu\nu}{}^a(P)\,.
\end{equation}
The Riemann curvature tensor is defined as
\begin{equation}\label{eq:Riemann}
\left[\overset{c}{\nabla}_\mu,\overset{c}{\nabla}_\nu\right]X_\sigma=\overset{c}{R}_{\mu\nu\sigma}{}^\rho X_\rho-2\overset{c}{\Gamma}{}^\rho_{[\mu\nu]}\nabla_\rho X_\sigma\,.
\end{equation}
This means that $\overset{c}{R}_{\mu\nu\sigma}{}^\rho$ is given by
\begin{equation}
\overset{c}{R}_{\mu\nu\sigma}{}^\rho=-\partial_\mu\overset{c}{\Gamma}{}^\rho_{\nu\sigma}+\partial_\nu\overset{c}{\Gamma}{}^\rho_{\mu\sigma}-\overset{c}{\Gamma}{}^\rho_{\mu\lambda}\overset{c}{\Gamma}{}^\lambda_{\nu\sigma}+\overset{c}{\Gamma}{}^\rho_{\nu\lambda}\overset{c}{\Gamma}{}^\lambda_{\mu\sigma}\,.
\end{equation}
By employing the vielbein postulates it can be shown that
\begin{equation}\label{eq:Riemothercurvs}
\overset{c}{R}_{\mu\nu\sigma}{}^\rho=-v^\rho e_{\sigma a}R_{\mu\nu}{}^a(C)-e_{\sigma a}e^\rho_b R_{\mu\nu}{}^{ab}(J)\,.
\end{equation}

We will now introduce a contravariant vector $M^\mu$ so as to realize the algebra of $\delta$ transformations \eqref{eq:deltatrafo1}--\eqref{eq:deltatrafo4} in terms of transformations of the smaller number of fields $\tau_\mu, e_\mu^a$ and $M^\mu$ (smaller as compared to the adjoint representation used earlier with the connections $\Omega_\mu{}^a$ and $\Omega_\mu{}^{ab}$). This can be achieved by taking $M^\mu$ to transform as
\begin{equation}
\delta M^\mu = \mathcal{L}_\xi M^\mu+\lambda^a e_a^\mu \,,
\end{equation}
and to find an expression for the affine connection in terms of $\tau_\mu, e_\mu^a$ and $M^\mu$ that is invariant under the local tangent space transformations. The vielbein postulates then make the connections $\Omega_\mu{}^a$ and $\Omega_\mu{}^{ab}$ fully expressable in terms of $\tau_\mu, e_\mu^a$ and $M^\mu$.

Using the field $M^\mu$ we can construct the following tangent space invariant objects
\begin{eqnarray}
\hat\tau_\mu & = & \tau_\mu-h_{\mu\nu}M^\nu\,,\\
\bar h^{\mu\nu} & = & h^{\mu\nu}-M^\mu v^\nu-M^\nu v^\mu\,,\\
\bar\Phi & = & -\tau_\mu M^\mu+\frac{1}{2}h_{\mu\nu}M^\mu M^\nu\,,
\end{eqnarray}
where $h^{\mu\nu}=\delta^{ab}e^\mu_a e^\nu_b$. There also exists a set of Carroll boost invariant vielbeins denoted by $\hat\tau_\mu$, $e_\mu^a$ whose inverses are $v^\mu$ and $\hat e^\mu_a$ where
\begin{equation}
\hat e^\mu_a = e^\mu_a-M^\nu e_{\nu a}v^\mu\,.
\end{equation}
We define $\hat h^{\mu\nu}=\delta^{ab}\hat e^\mu_a\hat e^\nu_b$. 

In terms of these invariants the affine connection can be taken to be
\begin{equation}\label{eq:affineconnection}
\overset{c}{\Gamma}{}^\lambda_{\mu\rho}=-v^\lambda\partial_\mu\hat\tau_\rho+\frac{1}{2}\hat h^{\nu\lambda}\left(\partial_\mu h_{\rho\nu}+\partial_\rho h_{\mu\nu}-\partial_\nu h_{\mu\rho}\right)-\hat h^{\nu\lambda}\hat\tau_\rho K_{\mu\nu}\,,
\end{equation}
where $K_{\mu\nu}$ is the extrinsic curvature defined as
\begin{equation}
K_{\mu\nu}=-\frac{1}{2}\mathcal{L}_v h_{\mu\nu}\,.
\end{equation}
The torsion tensor is therefore
\begin{equation}
2\overset{c}{\Gamma}{}^\lambda_{[\mu\rho]}=-2v^\lambda\partial_{[\mu}\hat\tau_{\rho]}-2\hat h^{\nu\lambda}\hat\tau_{[\rho} K_{\mu]\nu}\,.
\end{equation}
The affine connection $\overset{c}{\Gamma}{}^\lambda_{\mu\rho}$ has the properties that
\begin{equation}
\overset{c}{\nabla}{}_\mu\hat\tau_\nu=\overset{c}{\nabla}{}_\mu h_{\nu\rho}=\overset{c}{\nabla}{}_\mu v^\nu=\overset{c}{\nabla}{}_{\mu}\hat h^{\nu\rho}=0\,.
\end{equation}
Just like in the case of the gauging of the Galilei algebra the metric compatible and tangent space invariant affine connection is not unique \cite{Bekaert:2014bwa,Hartong:2015wxa,Hartong:2015zia}. This is not a problem one just needs to specify which connection is being used. In this paper we will always work with the connection \eqref{eq:affineconnection}. The traces of \eqref{eq:affineconnection} are given by
\begin{equation}
\overset{c}{\Gamma}{}_{\rho\nu}^\rho=-\mathcal{L}_v\hat\tau_\nu+e^{-1}\partial_\nu e-\hat\tau_\nu K\,,
\end{equation}
and
\begin{equation}
\overset{c}{\Gamma}{}_{\nu\rho}^\rho=e^{-1}\partial_\nu e\,,
\end{equation}
where
\begin{equation}
e=\text{det}\,\left(\tau_\mu, e_\mu\right)\,,
\end{equation}
so that
\begin{equation}
2\overset{c}{\Gamma}{}_{[\rho\nu]}^\rho=-\mathcal{L}_v\hat\tau_\nu-\hat\tau_\nu K\,.
\end{equation}

\subsection{Two-dimensions}\label{subsec:2D}

In two or rather 1+1 dimensions the spatial rotations are no longer part of the Carroll algebra. Further, the Carroll algebra is now isomorphic to the 2-dimensional Galilei algebra (without the central extension giving rise to the Bargmann algebra). To see the isomorphism one just needs to interchange the roles of $H$ and $P$. That means that in 1+1 dimensions we are dealing with torsional Newton--Cartan geometry. However we keep writing things in a Carrollian manner as that is the more natural geometry and because that is the geometry one finds on the boundary at future null infinity in higher dimensions which eases comparison. 

In this subsection we collect a few special relations that are useful when dealing with 2-dimensional Carrollian geometries. First of all in 1+1 dimensions the epsilon tensor is
\begin{eqnarray}
\epsilon_{\mu\nu} & = & \hat\tau_\mu e_\nu-\hat\tau_\nu e_\mu\,,\\
\epsilon^{\mu\nu} & = & v^\mu\hat e^\nu-v^\nu\hat e^\mu\,.
\end{eqnarray}
This can be used to compute the inverse vielbeins. 

The extrinsic curvature tensor is pure trace
\begin{equation}
K_{\mu\nu}=K h_{\mu\nu}\,,
\end{equation}
where
\begin{equation}\label{eq:K}
K=-v^\mu\hat e^\nu\left(\partial_\mu e_\nu-\partial_\nu e_\mu\right)=-e^{-1}\partial_\mu\left(ev^\mu\right)\,.
\end{equation}
Similarly using the epsilon tensor one can show that
\begin{equation}
\hat e^\mu\mathcal{L}_v\hat\tau_\mu=e^{-1}\partial_\mu\left(e\hat e^\mu\right)\,.
\end{equation}

Using the vielbein postulate \eqref{eq:VP1} we find
\begin{equation}
\partial_\mu\tau_\nu-\Gamma_{\mu\nu}^\rho\tau_\rho=\Omega_\mu e_\nu=\overset{c}{\nabla}{}_\mu\tau_\nu=h_{\nu\rho}\overset{c}{\nabla}{}_\mu M^\rho=e_\nu e_\rho\overset{c}{\nabla}{}_\mu M^\sigma
\end{equation}
so that
\begin{equation}
\Omega_\mu=\partial_\mu\left(e_\rho M^\rho\right)\,,
\end{equation}
and thus the curvature associated with local Carroll boosts \eqref{eq:Ccurv} is zero because
\begin{equation}
R_{\mu\nu}(C)=\partial_\mu\Omega_\nu-\partial_\nu\Omega_\mu=0\,.
\end{equation}
This means that since we have \eqref{eq:Riemothercurvs} which in 1+1 dimensions reads
\begin{equation}
\overset{c}{R}{}_{\mu\nu\rho}{}^\sigma=-v^\sigma e_\mu R_{\rho\nu}(C)
\end{equation}
the geometry is flat, i.e.
\begin{equation}
\overset{c}{R}{}_{\mu\nu\rho}{}^\sigma=0\,.
\end{equation}
The torsion however is given by
\begin{equation}\label{eq:torsion}
2\hat e^\mu v^\nu\overset{c}{\Gamma}{}_{[\mu\nu]}^\rho=v^\rho\hat e^\mu\mathcal{L}_v\hat\tau_\mu+\hat e^\rho K\,,
\end{equation}
and so in 1+1 dimensions Carrollian geometry is flat but it has torsion.

\section{Expansions up to NLO}\label{sec:expuptoNLO}

In section \ref{subsec:LO+NLO} we first solve the bulk equations of motion at leading and NLO order. To this end we collect here some useful results regarding the expansion of the metric and Christoffel connection needed to solve the Einstein equations at LO and NLO. The main purpose of section \ref{subsec:LO+NLO} is to establish equations  \eqref{eq:rel1} and \eqref{eq:rel2}.

The expansion of the metric is given by
\begin{eqnarray}
g_{rr} & = & 2\bar\Phi r^{-2}+\mathcal{O}(r^{-3})\,,\\
g_{r\mu} & = & -\hat\tau_\mu+r^{-1}h_{(1)r\mu}+\mathcal{O}(r^{-2})\,,\\
g_{\mu\nu} & = & r^2 h_{\mu\nu}+rh_{(1)\mu\nu}+\mathcal{O}(1)\,,
\end{eqnarray}
while the expansion of the inverse metric reads
\begin{eqnarray}
g^{rr} & = & rH_{(1)}^{rr}+\mathcal{O}(1)\,,\\
g^{r\mu} & = & v^\mu+r^{-1}H_{(1)}^{r\mu}+\mathcal{O}(r^{-2})\,,\\
g^{\mu\nu} & = & r^{-2}\bar h^{\mu\nu}+r^{-3}H_{(1)}^{\mu\nu}+\mathcal{O}(r^{-4})\,,
\end{eqnarray}
where
\begin{eqnarray}
H_{(1)}^{rr} & = & -v^\mu v^\nu h_{(1)\mu\nu}\,,\\
H_{(1)}^{r\mu} & = & -v^\mu v^\nu h_{(1)r\nu}-\hat h^{\mu\nu}v^\rho h_{(1)\nu\rho}+2\bar\Phi v^\mu v^\nu v^\rho h_{(1)\nu\rho}\,,\\
h_{\mu\nu}H_{(1)}^{\mu\nu} & = & -\hat h^{\mu\nu}h_{(1)\mu\nu}\,.
\end{eqnarray}
Other components of $H_{(1)}^{\mu\nu}$ are not needed.

Using these results we can compute the following expansions of the Christoffel connections
\begin{eqnarray}
\Gamma^r_{rr} & = & r^{-2}\left(2\bar\Phi v^\mu v^\nu h_{(1)\mu\nu}-v^\mu h_{(1)r\mu}-v^\mu\partial_\mu\bar\Phi\right)+\mathcal{O}(r^{-3})\,,\\
\Gamma^r_{r\mu} & = & \frac{1}{2}v^\nu h_{(1)\mu\nu}+\frac{1}{2}\mathcal{L}_v\hat\tau_\mu-h_{\mu\nu}\hat h^{\nu\rho}v^\sigma h_{(1)\rho\sigma}+\mathcal{O}(r^{-1})\,,\\
\Gamma^r_{\mu\nu} & = & r^2\left(K h_{\mu\nu}+h_{\mu\nu}v^\rho v^\sigma h_{(1)\rho\sigma}\right)+\mathcal{O}(r)\,,\\
\Gamma^\mu_{rr} & = & -2r^{-3}\bar\Phi v^\mu+\mathcal{O}(r^{-4})\,,\\
\Gamma^\mu_{r\nu} & = & r^{-1}h_{\nu\rho}\hat h^{\mu\rho}+r^{-2}\Gamma_{(1)r\nu}^\mu+\mathcal{O}(r^{-3})\,,\\
\Gamma^\rho_{\mu\nu} & = & -rv^\rho h_{\mu\nu}+\Gamma_{(1)\mu\nu}^\rho+\mathcal{O}(r^{-1})\,,
\end{eqnarray}
where
\begin{eqnarray}
\Gamma_{(1)r\nu}^\mu & = & v^\mu\partial_\nu\bar\Phi-v^\mu e_\nu\hat\tau_\rho\hat e^\sigma\Gamma_{(1)r\sigma}^\rho+\bar\Phi v^\mu\hat\tau_\nu v^\rho v^\sigma h_{(1)\rho\sigma}\nonumber\\
&&-\frac{1}{2}\hat e^\mu\hat\tau_\nu\hat e^\rho v^\sigma h_{(1)\rho\sigma}+\frac{1}{2}\hat e^\mu\hat\tau_\nu\hat e^\rho\mathcal{L}_v\hat\tau_\rho-\frac{1}{2}\hat e^\mu e_\nu\hat h^{\rho\sigma}h_{(1)\rho\sigma}\,,\\
\Gamma_{(1)\mu\nu}^\rho & = & \frac{1}{2}\overset{c}{\Gamma}{}_{(\mu\nu)}^\rho-\frac{1}{2}v^\rho h_{(1)\mu\nu}+\frac{1}{2}\hat h^{\rho\sigma}h_{\mu\sigma}\hat\tau_\nu K+\frac{1}{2}\hat h^{\rho\sigma}h_{\nu\sigma}\hat\tau_\mu
K-2\bar\Phi v^\rho h_{\mu\nu}K\nonumber\\
&&+h_{\mu\nu}\left(v^\rho v^\sigma h_{(1)r\sigma}+\hat h^{\rho\sigma}v^\lambda h_{(1)\sigma\lambda}-2\bar\Phi v^\rho v^\sigma v^\lambda h_{(1)\sigma\lambda}\right)\,.
\end{eqnarray}
The term $\hat\tau_\rho\hat e^\sigma\Gamma_{(1)r\sigma}^\rho$ is not needed.

\section{Asymptotic Expansions}\label{app:asymptotics}

In this last appendix we collect a large set of near boundary expansions of the metric, Christoffel connections and all the vielbeins. Further we write down the asymptotic expansions of the diffeomorphisms that leave invariant the BMS gauge with sources. Throughout this appendix we use equations \eqref{eq:rel1} and \eqref{eq:rel2} for $v^\mu h_{(1)r\mu}$ and $v^\mu h_{(1)\mu\nu}$.

\subsection{Metric}\label{subsec:metricexp}

For our purposes of solving the Einstein equations up to the order that provides us with the Ward identities and for computing the variation of the on-shell action it will prove sufficient to expand the metric up to the following orders
\begin{eqnarray}
g_{rr} & = & 2\bar\Phi r^{-2}+\mathcal{O}(r^{-3})\,,\label{eq:BMSsources1app}\\
g_{r\mu} & = & -\hat\tau_\mu+r^{-1}h_{(1)r\mu}+\mathcal{O}(r^{-2})\,,\\
g_{\mu\nu} & = & r^2h_{\mu\nu}+rh_{(1)\mu\nu}+h_{(2)\mu\nu}+\mathcal{O}(r^{-1})\,.\label{eq:BMSsources3app}
\end{eqnarray}
The expansion of the inverse metric is given by
\begin{eqnarray}
g^{rr} & = & rH_{(1)}^{rr}+H_{(2)}^{rr}+\mathcal{O}(r^{-1})\,,\label{eq:invgrr}\\
g^{r\mu} & = & v^\mu+r^{-1}H_{(1)}^{r\mu}+r^{-2}H_{(2)}^{r\mu}+\mathcal{O}(r^{-3})\,,\\
g^{\mu\nu} & = & r^{-2}\bar h^{\mu\nu}+r^{-3}H_{(1)}^{\mu\nu}+r^{-4}H_{(2)}^{\mu\nu}+\mathcal{O}(r^{-5})\,,
\end{eqnarray}
where
\begin{eqnarray}
H_{(1)}^{rr} & = & 2K\,,\label{eq;H1rr}\\
H_{(1)}^{r\mu} & = & -2\bar\Phi Kv^\mu-\hat h^{\mu\nu}\mathcal{L}_v\hat\tau_\nu+v^\mu v^\nu\partial_\nu\bar\Phi\,,\\
e_\mu\hat\tau_\nu H_{(1)}^{\mu\nu} & = & -2\bar\Phi\hat e^\mu\mathcal{L}_v\hat\tau_\mu+\hat e^\mu h_{(1)r\mu}\,,\\
e_\mu e_\nu H_{(1)}^{\mu\nu} & = & -\hat h^{\mu\nu}h_{(1)\mu\nu}\,,
\end{eqnarray}
we will not need $\hat\tau_\mu\hat\tau_\nu H_{(1)}^{\mu\nu}$, and where
\begin{eqnarray}
H_{(2)}^{rr} & = & -v^\mu v^\nu h_{(2)\mu\nu}+4Kv^\mu\partial_\mu\bar\Phi+\left(\hat e^\mu\mathcal{L}_v\hat\tau_\mu\right)^2\,,\label{eq:H2rr}\\
e_\mu H_{(2)}^{r\mu} & = & -2K\hat e^\mu h_{(1)r\mu}+\left(\hat e^\mu\mathcal{L}_v\hat\tau_\mu\right)\hat h^{\rho\sigma}h_{(1)\rho\sigma}-\left(\hat e^\mu\mathcal{L}_v\hat\tau_\mu\right)v^\nu\partial_\nu\bar\Phi\nonumber\\
&&+2\bar\Phi K\hat e^\mu\mathcal{L}_v\hat\tau_\mu-\hat e^\mu v^\nu h_{(2)\mu\nu}\,.
\end{eqnarray}
The components $\hat\tau_\mu H_{(2)}^{r\mu}$ and $H_{(2)}^{\mu\nu}$ will also not be needed.

\subsection{Connection}\label{subsec:connections}

We expand the connections up to the following orders
\begin{eqnarray}
\Gamma^r_{rr} & = & -2r^{-2}\bar\Phi K+\mathcal{O}(r^{-3})\,,\\
\Gamma^r_{r\mu} & = & \hat\tau_\mu K+r^{-1}\Gamma^r_{(2)r\mu}+\mathcal{O}(r^{-2})\,,\\
\Gamma^r_{\mu\nu} & = & -r^2Kh_{\mu\nu}+r\Gamma^r_{(2)\mu\nu}+\Gamma^r_{(3)\mu\nu}+\mathcal{O}(r^{-1})\,,\\
\Gamma^\mu_{rr} & = & -2r^{-3}\bar\Phi v^\mu+r^{-4}\Gamma^\mu_{(1)rr}+\mathcal{O}(r^{-5})\,,\\
\Gamma^\mu_{r\nu} & = & r^{-1}h_{\mu\rho}\hat h^{\nu\rho}+r^{-2}\Gamma^\mu_{(1)r\nu}+r^{-3}\Gamma^\mu_{(2)r\nu}+\mathcal{O}(r^{-4})\,,\\
\Gamma^\rho_{\mu\nu} & = & -rv^\rho h_{\mu\nu}+\Gamma^\rho_{(1)\mu\nu}+r^{-1}\Gamma^\rho_{(2)\mu\nu}+r^{-2}\Gamma^\rho_{(3)\mu\nu}+\mathcal{O}(r^{-3})\,,
\end{eqnarray}
where the first order coefficients are given by
\begin{eqnarray}
e_\mu\Gamma^\mu_{(1)rr} & = & -\hat e^\mu h_{(1)r\mu}-\hat e^\mu\partial_\mu\bar\Phi+2\bar\Phi\hat e^\mu\mathcal{L}_v\hat\tau_\mu\,,\\
\Gamma^\mu_{(1)r\nu} & = & v^\mu\partial_\nu\bar\Phi-v^\mu e_\nu\hat e^\rho h_{(1)r\rho}-\frac{1}{2}\hat h^{\mu\rho}h_{\rho\nu}\hat h^{\lambda\kappa}h_{(1)\lambda\kappa}-2\bar\Phi\hat\tau_\nu v^\mu K\,,\\
\Gamma^\rho_{(1)\mu\nu} & = & \overset{c}{\Gamma}{}_{(\mu\nu)}^\rho-v^\rho h_{\mu\nu}v^\rho\partial_\rho\bar\Phi+h_{\mu\nu}\hat h^{\rho\sigma}\mathcal{L}_v\hat\tau_\sigma+2v^\rho\hat\tau_\mu\hat\tau_\nu K-\frac{1}{2}v^\rho h_{\mu\nu}\hat h^{\lambda\kappa}h_{(1)\lambda\kappa}\nonumber\\
&&+\frac{1}{2}\left(\delta^\rho_\mu\hat\tau_\nu+\delta^\rho_\nu\hat\tau_\mu\right)K+\frac{1}{2}v^\rho\left(\hat\tau_\mu\mathcal{L}_v\hat\tau_\nu+\hat\tau_\nu\mathcal{L}_v\hat\tau_\mu\right)\,.
\end{eqnarray}
We will not be needing the component $\hat\tau_\mu\Gamma^\mu_{(1)rr}$. At second order in derivatives we have
\begin{eqnarray}
v^\mu\Gamma^r_{(2)r\mu} & = & Kv^\mu\partial_\mu\bar\Phi+2\bar\Phi K^2\,,\\
\hat e^\mu\Gamma^r_{(2)r\mu} & = & -v^\mu\hat e^\nu h_{(2)\mu\nu}-\frac{1}{2}\hat e^\mu\partial_\mu\left(v^\nu\partial_\nu\bar\Phi\right)-\frac{1}{2}v^\mu\partial_\mu\left(\hat e^\nu h_{(1)r\nu}\right)+\left(\hat e^\mu\partial_\mu\bar\Phi\right)K\nonumber\\
&&-\bar\Phi\hat e^\mu\partial_\mu K-\bar\Phi K\hat e^\mu\mathcal{L}_v\hat\tau_\mu-\frac{1}{2}\left(\hat e^\mu\mathcal{L}_v\hat\tau_\mu\right)v^\nu\partial_\nu\bar\Phi\nonumber\\
&&+\frac{1}{2}\left(\hat e^\mu\mathcal{L}_v\hat\tau_\mu\right)\hat h^{\lambda\kappa}h_{(1)\lambda\kappa}-\frac{3}{2}\hat e^\mu h_{(1)r\mu}K\,,\\
v^\mu v^\nu\Gamma^r_{(2)\mu\nu} & = & -v^\mu\partial_\mu K+2K^2\,,\\
v^\mu \hat e^\nu\Gamma^r_{(2)\mu\nu} & = & -\hat e^\mu\partial_\mu K-K\hat e^\mu\mathcal{L}_v\hat\tau_\mu\,,\\
\hat e^\mu\hat e^\nu \Gamma^r_{(2)\mu\nu} & = & \hat e^\mu\partial_\mu\left(\hat e^\nu\mathcal{L}_v\hat\tau_\nu\right)-\frac{1}{2}v^\mu\partial_\mu\left(\hat h^{\lambda\kappa}h_{(1)\lambda\kappa}\right)+\left(\hat e^\mu\mathcal{L}_v\hat\tau_\mu\right)^2\nonumber\\
&&+Kv^\mu\partial_\mu\bar\Phi-2\bar\Phi K^2\,.
\end{eqnarray}
Regarding $\Gamma^\mu_{(2)r\nu}$ and $\Gamma^\rho_{(2)\mu\nu}$ we will only need the following tangent space components
\begin{eqnarray}
e_\mu v^\nu\Gamma^\mu_{(2)r\nu} & = & -\frac{1}{2}\left(\hat e^\mu\mathcal{L}_v\hat\tau_\mu\right)v^\nu\partial_\nu\bar\Phi+\frac{1}{2}\hat e^\mu\partial_\mu\left(v^\nu\partial_\nu\bar\Phi\right)+K\hat e^\mu\partial_\mu\bar\Phi+\bar\Phi\hat e^\mu\partial_\mu K\nonumber\\
&&+\frac{1}{2}v^\mu\partial_\mu\left(\hat e^\nu h_{(1)r\nu}\right)-\bar\Phi K\hat e^\mu\mathcal{L}_v\hat\tau_\mu+\frac{1}{2}\hat e^\mu h_{(1)r\mu}K\,,\\
e_\rho v^\mu v^\nu\Gamma_{(2)\mu\nu}^\rho & = & \hat e^\mu\partial_\mu K+v^\mu\partial_\mu\left(\hat e^\nu\mathcal{L}_v\hat\tau_\nu\right)\,,\\
e_\rho v^\mu \hat e^\nu\Gamma_{(2)\mu\nu}^\rho & = & \frac{1}{2}v^\mu\partial_\mu\left(\hat h^{\rho\sigma}h_{(1)\rho\sigma}\right)\,,\\
e_\rho \hat e^\mu \hat e^\nu\Gamma^\rho_{(2)\mu\nu} & = & -\frac{1}{2}\left(\hat e^\mu\mathcal{L}_v\hat\tau_\mu\right)\hat h^{\rho\sigma}h_{(1)\rho\sigma}+\frac{1}{2}\hat e^\mu\partial_\mu\left(\hat h^{\rho\sigma}h_{(1)\rho\sigma}\right)+K\hat e^\mu h_{(1)r\mu}\nonumber\\
&&+\left(\hat e^\mu\mathcal{L}_v\hat\tau_\mu\right)v^\nu\partial_\nu\bar\Phi+\hat e^\mu v^\nu h_{(2)\mu\nu}\,,\\
\hat\tau_\rho v^\mu v^\nu\Gamma_{(2)\mu\nu}^\rho & = & 2\bar\Phi K^2 + Kv^\mu\partial_\mu\bar\Phi + v^\mu\partial_\mu\left(v^\nu\partial_\nu\bar\Phi\right)\,,\\
\hat\tau_\rho\hat e^\mu v^\nu\Gamma^\rho_{(2)\mu\nu} & = & -\bar\Phi K\hat e^\mu\mathcal{L}_v\hat\tau_\mu-\frac{1}{2}\hat e^\mu h_{(1)r\mu}K-\bar\Phi\hat e^\mu\partial_\mu K+K\hat e^\mu\partial_\mu\bar\Phi\nonumber\\
&&+\frac{1}{2}\hat e^\mu\partial_\mu\left(v^\nu\partial_\nu\bar\Phi\right)+\frac{1}{2}\left(e^\mu\mathcal{L}_v\hat\tau_\mu\right)v^\nu\partial_\nu\bar\Phi-\frac{1}{2}v^\mu\partial_\mu\left(\hat e^\nu h_{(1)r\nu}\right)\,.
\end{eqnarray}
The components $\hat\tau_\rho\hat e^\mu\hat e^\nu\Gamma_{(2)\mu\nu}^\rho$ and $\Gamma^r_{(2)rr}$ will not be needed. At third order in derivatives we only need the following two expressions
\begin{eqnarray}
v^\mu v^\nu\Gamma^r_{(3)\mu\nu} & = & -2Kv^\mu\partial_\mu\left(v^\nu\partial_\nu\bar\Phi\right)-Kv^\mu v^\nu h_{(2)\mu\nu}+\frac{1}{2}v^\rho\partial_\rho\left(v^\mu v^\nu h_{(2)\mu\nu}\right)\nonumber\\
&&-2\bar\Phi K v^\mu\partial_\mu K-\left(v^\mu\partial_\mu\bar\Phi\right)v^\nu\partial_\nu K-\left(\hat e^\mu\mathcal{L}_v\hat\tau_\mu\right)v^\rho\partial_\rho\left(\hat e^\nu\mathcal{L}_v\hat\tau_\nu\right)\nonumber\\
&&-\left(\hat e^\mu\mathcal{L}_v\hat\tau_\mu\right)\hat e^\nu\partial_\nu K\,,\\
v^\mu v^\nu e_\rho\Gamma^\rho_{(3)\mu\nu} & = & \left(\hat e^\mu\mathcal{L}_v\hat\tau_\mu\right)v^\rho\partial_\rho\left(v^\nu\partial_\nu\bar\Phi\right)+K\left(\hat e^\mu\mathcal{L}_v\hat\tau_\mu\right)v^\nu\partial_\nu\bar\Phi-2K^2\hat e^\mu h_{(1)r\mu}\nonumber\\
&&+2\bar\Phi K^2\hat e^\mu\mathcal{L}_v\hat\tau_\mu-2K\hat e^\mu v^\nu h_{(2)\mu\nu}+v^\rho\partial_\rho\left(\hat e^\mu v^\nu h_{(2)\mu\nu}\right)\nonumber\\
&&-\frac{1}{2}\hat e^\rho\partial_\rho\left(v^\mu v^\nu h_{(2)\mu\nu}\right)-\left(\hat e^\rho\mathcal{L}_v\hat\tau_\rho\right)v^\mu v^\nu h_{(2)\mu\nu}+\hat e^\mu h_{(1)r\mu}v^\nu\partial_\nu K\nonumber\\
&&-\hat h^{\lambda\kappa}h_{(1)\lambda\kappa}v^\mu\partial_\mu\left(\hat e^\nu\mathcal{L}_v\hat\tau_\nu\right)-\hat h^{\lambda\kappa}h_{(1)\lambda\kappa}\hat e^\mu\partial_\mu K\,.
\end{eqnarray}
Useful expressions for the expansion of the trace, $\Gamma^\rho_{\rho\mu}$, i.e. the derivative of $\log\sqrt{-g}$ are
\begin{eqnarray}
\Gamma^\rho_{(1)\rho\mu}+\Gamma^r_{(1)r\mu} & = & e^{-1}\partial_\mu e\,,\\
\Gamma^\rho_{(2)\rho\mu}+\Gamma^r_{(2)r\mu} & = & \partial_\mu\left(-v^\rho\partial_\rho\bar\Phi+\frac{1}{2}\hat h^{\rho\sigma}h_{(1)\rho\sigma}\right)\,.
\end{eqnarray}

\subsection{Vielbeins}\label{subsec:vielbeins}

For various calculations, notably those of sections \ref{subsec:normal}, \ref{subsec:intmeasure} and \ref{subsec:GHbdryterm}, it is necessary to know the near boundary expansion of the bulk vielbeins $U$, $V$, $E$ that are such that the bulk metric can be written as
\begin{equation}
ds^2=-2UV+EE\,,
\end{equation}
where $U$ and $V$ are null vectors normalized such that $g^{MN}U_MV_N=-1$ and where $E$ is a unit spacelike vector orthogonal to $U$ and $V$. 

The bulk vielbeins can be Taylor expanded as follows
\begin{eqnarray}
U_r & = & 1+r^{-1}U_{(1)r}+\mathcal{O}(r^{-2})\,,\\
U_\mu & = & rU_{(1)\mu}+U_{(2)\mu}+\mathcal{O}(r^{-1})\,,\\
V_r & = & r^{-2}\tau_\mu M^\mu+\mathcal{O}(r^{-3})\,,\\
V_\mu & = & \tau_\mu+r^{-1}V_{(1)\mu}+\mathcal{O}(r^{-2})\,,\\
E_r & = & r^{-1}e_\nu M^\nu+r^{-2}E_{(1)r}+\mathcal{O}(r^{-3})\,,\\
E_\mu & = & re_\mu+e_{(1)\mu}+r^{-1}e_{(2)\mu}+\mathcal{O}(r^{-2})\,.
\end{eqnarray}
The Taylor expansion of the vielbeins is chosen such that we reproduce the near boundary Taylor expansion of the metric using $ds^2=-2UV+EE$. Comparing with the metric expansion we find
\begin{eqnarray}
v^\mu e_{(1)\mu} & = & \hat e^\mu\mathcal{L}_v\hat\tau_\mu-Ke_\mu M^\mu-\hat e^\mu U_{(1)\mu}\,,\\
\hat e^\mu e_{(1)\mu} & = & \frac{1}{2}\hat h^{\mu\nu}h_{(1)\mu\nu}+e_\nu M^\nu \hat e^\mu U_{(1)\mu}\,,\\
v^\mu V_{(1)\mu} & = & v^\mu\partial_\mu\bar\Phi+2\bar\Phi K+U_{(1)r}+K\tau_\mu M^\mu+e_\nu M^\nu\hat e^\mu\mathcal{L}_v\hat\tau_\mu\nonumber\\
&&-K\left(e_\mu M^\mu\right)^2-e_\mu M^\mu\hat e^\nu U_{(1)\nu}\,,\\
\hat e^\mu V_{(1)\mu} & = & -\hat e^\mu h_{(1)r\mu}-e_\mu M^\mu U_{(1)r}-\tau_\mu M^\mu\hat e^\nu U_{(1)\nu}+E_{(1)r}\nonumber\\
&&+\left(e_\mu M^\mu\right)^2\hat e^\nu U_{(1)\nu}+\frac{1}{2}e_\rho M^\rho\hat h^{\mu\nu}h_{(1)\mu\nu}\,,\\
v^\mu e_{(2)\mu} & = & v^\mu\hat e^\nu h_{(2)\mu\nu}-K\hat e^\mu V_{(1)\mu}+\hat e^\nu U_{(1)\nu}v^\mu V_{(1)\mu}+v^\mu U_{(2)\mu}e_\nu M^\nu\nonumber\\
&&-\hat e^\mu U_{(2)\mu}-v^\mu e_{(1)\mu}\hat e^\nu e_{(1)\nu}\,.
\end{eqnarray}
The fact that $U_M$ must be null at first and second order leads to the conditions
\begin{eqnarray}
v^\mu U_{(1)\mu} & = & -K\,,\\
v^\mu U_{(2)\mu} & = & \frac{1}{2}v^\mu v^\nu h_{(2)\mu\nu}-\bar\Phi K^2-Kv^\mu\partial_\mu\bar\Phi-KU_{(1)r}\nonumber\\
&&-\frac{1}{2}\left(\hat e^\mu U_{(1)\mu}-\hat e^\mu\mathcal{L}_v\hat\tau_\mu\right)^2\,.\label{eq:vU2}
\end{eqnarray}
The inverse vielbeins are expanded as follows
\begin{eqnarray}
U^r & = & rK+U_{(2)}^r+\mathcal{O}(r^{-1})\,,\\
U^\mu & = & v^\mu+r^{-1}U_{(1)}^\mu+\mathcal{O}(r^{-2})\,,\\
V^r & = & -1+r^{-1}V_{(1)}^r+\mathcal{O}(r^{-2})\,,\\
V^\mu & = & r^{-2}M^\mu+\mathcal{O}(r^{-3})\,,\\
E^r & = & E^r_{(1)}+\mathcal{O}(r^{-1})\,,\\
E^\mu & = & r^{-1}e^\mu+\mathcal{O}(r^{-2})\,,
\end{eqnarray}
where
\begin{eqnarray}
U_{(2)}^r & = & -v^\mu U_{(2)\mu}-\left(\hat e^\mu U_{(1)\mu}\right)^2+\left(\hat e^\mu\mathcal{L}_v\hat\tau_\mu\right)\hat e^\nu U_{(1)\nu}+Kv^\mu\partial_\mu\bar\Phi\,,\\
U_{(1)}^\mu & = & v^\mu U_{(1)r}+v^\mu v^\nu\partial_\nu\bar\Phi-\hat e^\mu\hat e^\nu\mathcal{L}_v\hat\tau_\nu+\hat e^\mu\hat e^\nu U_{(1)\nu}\,,\\
V_{(1)}^r & = & U_{(1)r}+2\bar\Phi K+K\tau_\mu M^\mu-e_\mu M^\mu\hat e^\nu U_{(1)\nu}\,,\\
E_{(1)}^r & = & Ke_\mu M^\mu-\hat e^\mu U_{(1)\mu}\,.
\end{eqnarray}

\subsection{Asymptotic diffeomorphisms}\label{subsec:asymptdiffs}

For many purposes such as finding appropriate metric parameterizations such as in section \ref{subsec:flatPBHtrafos} where we write down the most general solution with a torsion free boundary, or for the computation of BMS symmetries which are all diffeomorphisms leaving invariant the boundary sources for a torsion free boundary \ref{subsec:BMS}, or concerning questions about the properties of the normal vector as in sections \ref{subsec:PBHtrafos} and \ref{subsec:GHbdryterm} it is important to have the bulk diffeomorphisms available that leave the BMS gauge with sources, \eqref{eq:BMSsources1app}--\eqref{eq:BMSsources3app}, form invariant. 

We expand the bulk generator of diffeomorphisms $\xi^M$ as follows
\begin{eqnarray}
\xi^r & = & r\Lambda_D+\xi^r_{(1)}+r^{-1}\xi^r_{(2)}+\mathcal{O}(r^{-2})\,,\label{eq:xir}\\
\xi^\mu & = & \chi^\mu+r^{-1}\chi^\mu_{(1)}+r^{-2}\chi^\mu_{(2)}+\mathcal{O}(r^{-3})\,.\label{eq:ximu}
\end{eqnarray}
By acting on the metric with such a diffeomorphism, 
\begin{eqnarray}
\mathcal{L}_\xi g_{rr} & = & 2r^{-2}\delta\bar\Phi+\mathcal{O}(r^{-3})\,,\\
\mathcal{L}_\xi g_{r\mu} & = & -\delta\hat\tau_\mu+r^{-1}\delta h_{(1)r\mu}+\mathcal{O}(r^{-2})\,,\\
\mathcal{L}_\xi g_{\mu\nu} & = & r^2\delta h_{\mu\nu}+r\delta h_{(1)\mu\nu}+\delta h_{(2)\mu\nu}+\mathcal{O}(r^{-1})\,,
\end{eqnarray}
we can read off the transformation of the sources
\begin{eqnarray}
\delta\bar\Phi & = & \mathcal{L}_\chi\bar\Phi+\hat\tau_\mu\chi^\mu_{(1)}\,,\label{eqLdiffeobarPhi}\\
\delta\hat\tau_\mu & = & \Lambda_D\hat\tau_\mu+\mathcal{L}_\chi\hat\tau_\mu+h_{\mu\rho}\chi_{(1)}^\rho\,,\\
\delta e_\mu & = & \Lambda_D e_\mu+\mathcal{L}_\chi e_\mu\,.
\end{eqnarray}
The inverse vielbeins transform as
\begin{eqnarray}
\delta v^\mu & = & -\Lambda_D v^\mu+\mathcal{L}_\chi v^\mu\,,\label{eq:sourcetrafo1}\\
\delta\hat e^\mu & = & -\lambda_D\hat e^\mu+\mathcal{L}_\chi\hat e^\mu+v^\mu e_\nu\chi_{(1)}^\nu\,.\label{eq:sourcetrafo2}
\end{eqnarray}

Next we take a look at the subleading components in the metric expansion. They transform as
\begin{eqnarray}
\delta h_{(1)rr} & = & -\Lambda_D h_{(1)rr}-4\bar\Phi\xi_{(1)}^r+\mathcal{L}_\chi h_{(1)rr}+2\mathcal{L}_{\chi_{(1)}}\bar\Phi-2h_{(1)r\mu}\chi_{(1)}^\mu+4\hat\tau_\mu\chi_{(2)}^\mu\,,\label{eq:h1rr}\\
\delta h_{(1)r\mu} & = & \mathcal{L}_\chi h_{(1)r\mu}-\mathcal{L}_{\chi_{(1)}}\hat\tau_\mu+2\bar\Phi\partial_\mu\Lambda_D-h_{(1)\mu\nu}\chi^\nu_{(1)}-2h_{\mu\nu}\chi_{(2)}^\nu\,,\\
\delta h_{(1)\mu\nu} & = & \Lambda_D h_{(1)\mu\nu}+2\xi^r_{(1)}h_{\mu\nu}+\mathcal{L}_\chi h_{(1)\mu\nu}+\mathcal{L}_{\chi_{(1)}}h_{\mu\nu}-\hat\tau_\mu\partial_\nu\Lambda_D-\hat\tau_\nu\partial_\mu\Lambda_D\,,\\
\delta h_{(2)\mu\nu} & = & \xi_{(1)}^r h_{(1)\mu\nu}+2\xi^r_{(2)}h_{\mu\nu}+\mathcal{L}_\chi h_{(2)\mu\nu}+\mathcal{L}_{\chi_{(1)}} h_{(1)\mu\nu}+\mathcal{L}_{\chi_{(2)}} h_{\mu\nu}\nonumber\\
&&-\hat\tau_\mu\partial_\nu\xi^r_{(1)}-\hat\tau_\nu\partial_\mu\xi^r_{(1)}+h_{(1)r\mu}\partial_\nu\Lambda_D+h_{(1)r\nu}\partial_\mu\Lambda_D\,.
\end{eqnarray}
We included the transformation of $h_{(1)rr}=0$, which appears subleading to $\bar\Phi$ in the expansion of $g_{rr}$ at order $r^{-3}$, i.e. 
\begin{equation}
g_{rr}=2\bar\Phi r^{-2}+r^{-3}h_{(1)rr}+\mathcal{O}(r^{-4})\,.
\end{equation}

Regarding the transformation of $U_M$ we consider both bulk diffeomorphisms and local Lorentz boosts that rescale $U_M$ by a local function. The reason for this is that we used the latter to fix $U_r$ to be equal to one at leading order so when performing local dilatations induced by a bulk diffeomorphism we need to perform a compensating local Lorentz boost to keep $U_r$ equal to unity at leading order. Hence we consider
\begin{equation}\label{eq:trafoUM}
\delta U_M=\mathcal{L}_\xi U_M+\bar\lambda U_M\,,
\end{equation}
where 
\begin{equation}
\bar\lambda=\alpha+r^{-1}\alpha_{(1)}+\mathcal{O}(r^{-2})\,.
\end{equation}
If we take the $r$ component and demand that the right hand side of \eqref{eq:trafoUM} is order $r^{-1}$ we need that
\begin{equation}
\alpha=-\Lambda_D\,.
\end{equation}
Using this we find the following transformations of the normal vector
\begin{eqnarray}
\delta U_{(1)r} & = & -\Lambda_D U_{(1)r}+ \mathcal{L}_\chi U_{(1)r}-U_{(1)\mu}\chi_{(1)}^\mu+\alpha_{(1)}\,,\label{eq:trafoU1r}\\
\delta U_{(1)\mu} & = & \partial_\mu\Lambda_D+\mathcal{L}_\chi U_{(1)\mu}\,,\label{eq:trafoU1}\\
\delta U_{(2)\mu} & = & -\Lambda_D U_{(2)\mu}+\mathcal{L}_\chi U_{(2)\mu}+\left(\xi^r_{(1)}+\alpha_{(1)}\right)U_{(1)\mu}+\mathcal{L}_{\chi_{(1)}}U_{(1)\mu}\nonumber\\
&&+U_{(1)r}\partial_\mu\Lambda_D+\partial_\mu\xi_{(1)}^r\,.\label{eq:trafoU2}
\end{eqnarray}

\renewcommand{\theequation}{\thesection.\arabic{equation}}


\providecommand{\href}[2]{#2}\begingroup\raggedright\endgroup

\end{document}